\documentclass[seceq]{ptptex}

\usepackage{graphicx}

\newcommand{\be}{\begin{equation}}
\newcommand{\ee}{\end{equation}}
\newcommand{\bea}{\begin{eqnarray}}
\newcommand{\eea}{\end{eqnarray}}
\newcommand{\beann}{\begin{eqnarray*}}
\newcommand{\eeann}{\end{eqnarray*}}
\newcommand{\nn}{\nonumber}
\newcommand{\ba}{\begin{array}}
\newcommand{\ea}{\end{array}}

\newcommand{\Tr}{\mathop{\rm Tr}}

\newcommand{\diag}{\mathop{\rm diag}\nolimits}

\newcommand{\e}{\epsilon}
\newcommand{\Z}{\mathbb{Z}}

\newcommand{\C}{\mathbb{C}}

\newcommand{\del}{\partial}
\newcommand{\D}{{\cal D}}

\newcommand{\calV}{{\cal V}}

\newcommand{\Kahler}{K\"ahler }
\newcommand{\HK}{hyper-K\"ahler }
\newcommand{\zb}{{\bar{z}}}
\newcommand{\W}{{\cal W}}
\newcommand{\M}{{\cal M}}

\newcommand{\Vol}{{\rm Vol}}
\newcommand{\A}{{\tilde{\cal A}}}
\newcommand{\koumoku}[1]{\vspace{5pt}
\noindent{\underline{\it #1}}\vspace{3pt}}

\title{
Volume of Moduli Space of Vortex Equations and Localization %
}

\author{
Akiko \textsc{Miyake}$^{1,2,}$\footnote{E-mail: miyake@ippan.kushiro-ct.ac.jp},
Kazutoshi \textsc{Ohta}$^{2,}$\footnote{E-mail: kohta@law.meijigakuin.ac.jp}
and Norisuke \textsc{Sakai}$^{3,}$\footnote{E-mail: sakai@lab.twcu.ac.jp}
}

\inst{
$^1$Department of General Education, Kushiro National College of Technology,
Kushiro 084-0916, Japan\\
$^2$Institute of Physics, Meiji Gakuin University, 
Yokohama 244-8539, Japan\\
$^3$Department of Mathematics, Tokyo Woman's Christian University, 
Tokyo 167-8585, Japan
}

\abst{
We evaluate volume of moduli space of BPS vortices on a 
compact genus $h$ Riemann surface $\Sigma_h$ 
by using topological field theory and localization 
technique developed by Moore, Nekrasov and Shatashvili. 
We apply this technique to Abelian (ANO) vortex and show 
that the volume of moduli space agrees with the previous 
results obtained by integrating over the moduli space metric.
We extend the evaluation to non-Abelian gauge groups 
and multi-flavors. We also compare our results with the 
volume of the \Kahler quotient space inspired by the brane configuration.
}

\begin{document}

\maketitle

\section{Introduction}

At the critical coupling, static solitons exert no force 
between them and are called BPS solitons\cite{Manton:2004tk}. 
Consequently the solutions of the BPS equations have many 
parameters such as the position of the soliton, which are 
called moduli. 
The moduli space of the BPS solitons 
plays important roles in understanding dynamics of 
solitons such as the scattering 
problem\cite{Manton:1981mp,Samols:1991ne}. 
In order to study the properties of the solitons, 
it is desirable to know not only the topologies but also 
the details of the geometries of the moduli space, 
including the metric. 
However, important information on the dynamics of the 
BPS solitons can often be obtained by just knowing the 
volume of the moduli space, which is defined by an integral 
of the volume form made from the metric over the moduli 
space.

Straightforwardly, the thermodynamical partition function 
can be obtained from the volume of the moduli space, 
since the solitons behave as free particles on their 
moduli space, when they move slowly\cite{Manton:1993tt,Shah:1993us}. 
We can evaluate the thermodynamical properties 
of solitons like the equation of states. 
There is another non-trivial and important application of 
the volume of the moduli space, which has been found first 
in the case of instantons. 
Nekrasov pointed out that the volume of the moduli 
space of the instantons, which is the so-called 
Nekrasov partition function, gives the non-perturbative 
effective prepotential of ${\cal N}=2$ supersymmetric 
Yang-Mills theory.\cite{Nekrasov:2002qd}  
This means that the partition function of the 
supersymmetric Yang-Mills theory essentially evaluates 
the volume of the BPS solitons which cause the 
non-perturbative quantum corrections. 
Thus the volume of the moduli space is very important 
to understand the thermodynamics and non-perturbative 
dynamics of the BPS solitons, although it is just one 
of quantitative properties of the moduli space. 

We can define the metric and the volume form on the moduli 
space from the effective Lagrangian. Integration of the 
volume form over the whole moduli space gives the volume. 
This procedure may be called a direct approach, but is 
difficult in practice. 
Most of the previous works\cite{Taubes:1979tm,Manton:1998kq,
Nasir:1998kt,Manton:2002wb,Chen:2004xu,Romao:2005ph}
 based on effective Lagrangian 
considered Abelian gauge theory with single charged scalar 
field at the critical coupling, allowing the BPS vortices. 
This vortex is called Abrikosov-Nielsen-Olesen 
(ANO) vortices,\cite{ANO} or Abelian local vortices. 
Finite values of the volume of the moduli space 
can be obtained for vortices on compact base manifolds 
such as genus $h$ Riemann surfaces. 
One of the most interesting results on Riemann surfaces 
with the finite area ${\cal A}$ is the upper 
bound\cite{Bradlow:1990ir,Manton:2010sa,Manton:2009ja} for 
the number of BPS vortices $k$. 
It has also been found that full details of the metric 
are not needed and that only a certain local structure 
of the moduli space of the BPS solutions is important 
to evaluate the total volume of the moduli 
space\cite{Manton:1998kq}. 
In another approach\cite{Romao:2005ph} too, the metric on the whole 
moduli space is not required. It is replaced by a topological 
information using the Duistermaat-Heckman localisation formula. 
More recently, BPS vortices in non-Abelian gauge theories 
have attracted much attention.\cite{Hanany:2003hp,Eto:2005yh,
Eto:2006mz,Eto:2006pg,Eto:2006uw,Eto:2007yv,Eto:2007aw,
Baptista:2008ex,Eto:2009wq,Manton:2010wu} 
The asymptotic metric of the moduli space is obtained for 
well-separated of non-Abelian vortices.\cite{Fujimori:2010fk} 
However, it is not enough to obtain the volume of the moduli 
space of the non-Abelian vortex. 

Another indirect evaluation of the volume of the moduli 
space is developed by Moore, Nekrasov and 
Shatashvili.\cite{Moore:1997dj,Gerasimov:2006zt} 
They have not used the metric of the moduli space, 
but the ``localization'' technique.\cite{Witten:1992xu,Blau:1993tv} 
The moduli space of the BPS solitons is usually 
a \Kahler manifold. 
This \Kahler structure induces the localization 
property in the integration of the \Kahler form 
over the moduli space which gives the volumes. 
The localization 
means that the integral of the 
volume form over the moduli space is localized 
(dominated) at isometry fixed points of the moduli space. 
This localization simplifies 
the evaluation of the volume of the 
moduli space drastically: 
the volume of the moduli space has been evaluated 
in this way for the Hitchin equation, which is the 
two-dimensional vortex system coupled with 
the Higgs field in the adjoint 
representation.\cite{Moore:1997dj,Gerasimov:2006zt} 
The localization technique is based on a topological field 
theory, which can be understood as a twisted version 
of the supersymmetric theories.\cite{Witten:1992xu}

In this paper, we apply the localization technique 
in the evaluation of the moduli space of the BPS 
vortices on Riemann surfaces\footnote{
Although Ref.~\citen{Romao:2005ph} also used a 
localization theorem, it is applied in a quite different 
context, and the relation to the localization method used 
in this paper is unclear.
}. 
We consider $N_f$ flavors of Higgs fields in the 
fundamental representation 
in the Abelian as well as the non-Abelian $U(N_c)$ gauge theories. 
For the ANO vortices ($N_c=N_f=1$) where volumes were obtained 
from the metric before\cite{Shah:1993us,Manton:1998kq}, 
we find that our results by 
the localization technique completely agree with the 
previous results for any topology of the Riemann surface. 
Although it has been difficult to construct the metric 
of moduli space of the non-Abelian vortex apart from 
well-separated local vortices ($N_f=N_c$),\cite{Fujimori:2010fk} 
we can evaluate the volume of the moduli space of the 
non-Abelian vortices, since the localization technique 
does not need the details of the BPS solutions nor the 
metric. 
We only need the BPS equations as constraints in the field 
configuration space. 
The moduli space of BPS solitons can be regarded as 
the quotient space of the fields constrained by the BPS 
equations, in contrast to the usual \Kahler and 
hyper-\Kahler quotient spaces which are defined 
in terms of the spatial coordinates. 
We have to extend the coordinate integrals 
to path integrals of the fields. 
But this extension is straightforward and 
the localization properties still hold in the path integral.
We will show that the integrals, which give the volume 
of the moduli space in the localization technique, reduce 
to simple residue integrals. 
So we can evaluate the volume of the moduli space of 
the BPS vortices much easier than the explicit construction 
of the metric from the BPS solutions. 
We also work out the metric of the moduli space of 
single Abelian as well as non-Abelian vortices in 
order to compare it to our result from localization 
technique. 

In \S2, we first take up the volume of general \Kahler and 
hyper-\Kahler quotient spaces as examples in order to 
explain an outline of the localization technique 
by introducing ambient supermanifolds. 
In \S3, we apply the localization technique to the 
BPS Abelian vortices 
including the multi-flavor case ($N_f \ge 1$). 
We further apply our localization technique to the 
BPS vortices in the non-Abelian gauge theories in \S4. 
The results of our localization technique are made 
more explicit for different topologies of the 
Riemann surfaces in \S5. 
Effective Lagrangians of single vortices are worked out 
for Abelian as well as non-Abelian gauge theories in \S6. 
Localization technique is applied to the Hanany-Tong 
moduli space in \S7. 
Relation of our localization technique to supersymmetry 
is given in \S8. 
Further discussion is given in \S9. 

\section{The Volume of Quotient Spaces}

In this first section, we review a calculation of the 
volume of (hyper-)\Kahler quotient spaces with a $U(1)$ 
isometry by using the localization technique, 
following Ref.~\citen{Lee:2006ys}.
We here consider a quotient space in terms of the 
usual spatial coordinates, but essentials of the 
calculations can be applied to field theoretical 
quotient space of the moduli space in subsequent 
sections.

\subsection{\Kahler quotient space}

We first consider a $2n$-dimensional \Kahler manifold 
$\cal M$ whose (real) coordinates are denoted as $x^i$ 
($i=1,\ldots,2n$). 
The \Kahler manifold possesses a \Kahler 2-form 
$\Omega$ which gives a volume form on $\cal M$. 
Then the volume of the \Kahler manifold is given by 
\be
\Vol(\M) = \frac{1}{n!}\int_{\M} \Omega^n. 
\ee
Introducing Grassmann variables $\psi^i$, it can be also 
written by
\be
\Vol(\M) = \int \prod_{i=1}^{2n} dx^i d\psi^i 
e^{-\frac{1}{2}\Omega_{ij}\psi^i\psi^j},
\label{Kahler volume}
\ee
where $\Omega_{ij}$ are anti-symmetric components of $\Omega$.

The \Kahler manifold ${\cal M}$ is now defined as a 
quotient space in a $(2n{+}2)$-dimensional ambient space 
$\hat\M$
\be
\M = \frac{\mu^{-1}(0)}{U(1)},
\ee
where $\mu(x^i,y,\theta)$ is a moment map defined in $\hat\M$
and $\mu^{-1}(0)$ represents a space of solutions $\mu=0$.
The \Kahler quotient space is a space of solutions $\mu=0$ 
with $U(1)$ identification, 
and the coordinate $y$ represents a transverse coordinate 
to the surface $\mu=0$ 
and $\theta$ is a $U(1)$ direction in $\hat\M$.
Since $\mu$ and $\theta$ is constant on $\M$ by definition, 
we can insert identities
\be
\int d\mu \delta(\mu) = \frac{1}{2\pi}\int d\theta = 1
\ee
into the volume integral (\ref{Kahler volume}), then we have
\be
\Vol(\M) = \frac{1}{2\pi}
\int \prod_{i=1}^{2n} dx^i d\psi^i d\mu d\theta 
\, \delta(\mu)e^{-\frac{1}{2}\Omega_{ij}\psi^i\psi^j}.
\ee
We can also write
\be
\delta(\mu) = \frac{1}{2\pi}\int d\phi \, e^{-i\phi \mu}.
\ee
Thus in terms of an integral over the coordinates of 
$\hat\M$, the volume of $\M$ is given by
\be
\Vol(\M) = \frac{1}{(2\pi)^2}
\int d\phi \prod_{i=1}^{2n} dx^i 
d\psi^i dy d\theta d\chi d\tilde{\chi} \,
e^{-S},
\ee
where $S$ is an ``action''
\be
S=i\phi \mu+\frac{1}{2}\Omega_{ij}\psi^i\psi^j
+\chi\frac{\del\mu}{\del y} \tilde{\chi},
\label{Kahler action}
\ee
and we have introduced Grassmann coordinates $\chi$ and 
$\tilde{\chi}$ to exponentiate a Jacobian 
$\frac{\del\mu}{\del y}$ of a change of variables 
from $\mu$ to $y$.

We now introduce the $(2n{+}2)$ coordinates of 
$\hat\M$ by $x^{\mu}=(x^i,y,\theta)$ 
and $(2n{+}2)$  Grassmann coordinates by 
$\psi^\mu=(\psi^i,\tilde{\chi},\chi)$ ($\mu=1,\ldots,2n+2$). 
The moment map also generates the $U(1)$ isometry in 
$\hat\M$, that is,
\be
d\mu + i_V\hat\Omega=0,
\ee
where $\hat\Omega$ is \Kahler 2-form on $\hat\M$ and 
$i_V$ is an interior product with respect 
to a vector field $V$ generated by the $U(1)$ isometry. 
In the components, it means 
\be
\frac{\del \mu}{\del y} + \hat\Omega_{\theta y}=0.
\ee
So the action (\ref{Kahler action}) can be written in 
terms of $x^\mu$ and $\psi^\mu$ covariantly 
\be
S=i\phi \mu(x) + \frac{1}{2}\hat{\Omega}_{\mu\nu}\psi^\mu\psi^\nu.
\label{Kahler action 2}
\ee
Thus we finally obtain the integral of the volume of 
the \Kahler quotient space $\M$ 
\be
\Vol(\M) = \frac{1}{(2\pi)^2}
\int d\phi \prod_{\mu=1}^{2n+2} dx^\mu d\psi^\mu \,
e^{-S(x^\mu,\psi^\mu)}.
\label{integral of Kahler}
\ee

The volume of the \Kahler quotient space is now converted 
into an 
integral (\ref{integral of Kahler}) 
over bosonic and fermionic coordinates. 
The ``localization'' mechanism owing to a symmetry 
of the above system will be 
very useful to evaluate the integral. 
Indeed if we introduce the following fermionic 
symmetry (BRST symmetry)
\be
\begin{array}{lcl}
Qx^i = i\psi^i && Q \psi^i = 0,\\
Qy = i\tilde{\chi} && Q \tilde{\chi} = 0,\\
Q\theta = i\chi && Q \chi = -\phi,\\
Q \phi = 0, && \\
\end{array}
\ee
we find that the action (\ref{Kahler action}) is invariant 
under this symmetry, namely $QS=0$.
Or we can define covariantly in the coordinates of 
the ambient space (supermanifold), 
\be
\begin{array}{lcl}
Qx^\mu = i\psi^\mu && Q \psi^\mu = -\phi V^\mu,\\
Q \phi = 0, && \\
\label{Kahler BRST symmetry}
\end{array}
\ee
where $V^\mu$ is a vector along the isometry direction 
$\theta$, which satisfies $V^\theta =1$ and others are zero.
If we add a BRST exact action $S'\equiv Q\Xi(x^\mu,\psi^\mu)$ 
to this system, the volume does not change in the
integral (\ref{integral of Kahler}). 
In other words, the volume integral modified by $S'$
\be
\Vol(\M) = \frac{1}{(2\pi)^2}
\int d\phi \prod_{\mu=1}^{2n+2} dx^\mu d\psi^\mu \,
e^{-S(x^\mu,\psi^\mu)-\frac{1}{g^2}Q\Xi(x^\mu,\psi^\mu)}
\label{modified integral of Kahler}
\ee
is independent of the ``coupling'' $g$.
This fact means that the integral is exact in the 
WKB approximation $g\to \infty$.
If we suitably choose $S'$, we can generally show 
that supports of the integral (\ref{integral of Kahler})
as well as 
(\ref{modified integral of Kahler}) 
are localized at fixed point of the BRST symmetry 
(\ref{Kahler BRST symmetry}).
This localization property is the reason why we can 
evaluate the volume integral (\ref{integral of Kahler}),
and will play an important role in subsequent discussions.

\subsection{\HK quotient space}

Next we extend the above discussions to \HK quotient spaces. 
The \HK quotient space has $SU(2)$ isometry 
and is defined by zero solutions of three moment maps 
$\mu_a=0$ ($a=1,2,3$) with the $U(1)$ identification
\be
\M = \frac{\mu_1^{-1}(0)\cap \mu_2^{-1}(0)\cap \mu_3^{-1}(0)}{U(1)}.
\ee
The three moment maps consist  
of 
a triplet of the $SU(2)$ isometry.
We introduce coordinates of $\M$ by $x^i$ ($i=1,\ldots,2n$).
Similarly to the \Kahler quotient case, the volume of $\M$ 
is calculated by
\bea
\Vol(\M) &=& \int_\M \Omega^n\nonumber\\
&=& \int \prod_{i=1}^{2n} dx^i d\psi^i 
e^{-\frac{1}{2}\Omega_{ij}\psi^i\psi^j},
\label{HK volume}
\eea
where $\Omega$ is a \Kahler 2-form associated with the 
third moment map $\mu_3$.
Inserting the moment map constraints $\mu_a=0$ into 
the integral of volume (\ref{HK volume}), we obtain the 
integral over the $(2n+4)$-dimensional ambient space 
$\hat\M$ with the Grassmann coordinates $\psi^i$
\be
\Vol(\M) = \frac{1}{2\pi}\int \prod_{i=1}^{2n} 
\left(dx^i d\psi^i \right)
\prod_{a=1}^3 dy^a d\theta \det 
\left(\frac{\del \mu_a}{\del y^b}\right)
\prod_{a=1}^3 \delta(\mu_a) \,
e^{-\frac{1}{2}\Omega_{ij}\psi^i\psi^j}.
\ee
Introducing additional Grassmann coordinates 
$(\chi_a,\tilde{\chi}_a)$ ($a=1,2,3$) 
and Lagrange multipliers  $(\phi,Y_1,Y_2)$ for the 
delta functions, 
the volume integral can be written as 
\be
\Vol(\M) = \frac{1}{(2\pi)^4}\int d\phi dY_1 dY_2
\prod_{i=1}^{2n}  \left(dx^i d\psi^i \right)
\prod_{a=1}^3 \left(dy^a d\chi^a d\tilde{\chi}^a\right)
d\theta \, e^{-S},
\ee
where the ``action'' $S$ is defined by
\be
S = i Y_1 \mu_1 +  i Y_2 \mu_2 + i\phi \mu_3 
+ \frac{1}{2}\Omega_{ij}\psi^i\psi^j
+\hat{\Omega}_{ia}\psi^i\tilde{\chi}^a
+\frac{1}{2}\hat{\Omega}_{ab}\tilde{\chi}^a\tilde{\chi}^b
+ \sum_{a,b=1}^3 \chi^a \frac{\del \mu_a}{\del y^b} \tilde{\chi}^b,
\ee
where extra terms $\hat{\Omega}_{ia}\psi^i\tilde{\chi}^a$ 
and $\frac{1}{2}\hat{\Omega}_{ab}\tilde{\chi}^a\tilde{\chi}^b$  
do 
not change the fermionic determinants of $\Omega_{ij}$ 
and $\frac{\del{\mu_a}}{\del y^b}$ 
since they can be eliminated by shifting of $\psi^i$ 
and $\tilde{\chi}^a$. 
The third moment map determines the \Kahler 2-form on 
$\hat\M$ by
\be
\frac{\del \mu_3}{\del y^a} + \hat\Omega_{\theta a} = 0. 
\qquad (a=1,2,3)
\ee
If we introduce the $(2n+4)$-dimensional coordinates of 
$\hat\M$ as $x^\mu = (x^i,y^a, \theta)$ and 
corresponding Grassmann coordinates 
$\psi^\mu=(\psi^i,\tilde{\chi}^a,\chi_3)$,
the action can be written as
\be
S = i Y_1 \mu_1 +  i Y_2 \mu_2 + i\phi \mu_3 
+ \frac{1}{2}\hat{\Omega}_{\mu\nu}\psi^\mu\psi^\nu
+ \sum_{a=1}^3 \left\{
\chi^1 \frac{\del \mu_1}{\del y^a} \tilde{\chi}^a 
+ \chi^2 \frac{\del \mu_2}{\del y^a} \tilde{\chi}^a
\right\}.
\ee
Using this action, we finally find an integral 
expression of the volume of \HK quotient space ${\cal M}$
\be
\Vol(\M) = \frac{1}{(2\pi)^4}\int d\phi dY_1 d\chi_1 
dY_2 d\chi_2  \prod_{\mu=1}^{2n+4} dx^\mu d\psi^\mu
\, e^{-S},
\label{Volume of HK}
\ee
where the integral is taken over the $(2n+4)$-dimensional 
ambient supermanifold with the coordinates 
$(x^\mu,\psi^\mu)$ and Lagrange multipliers $(Y_1,\chi_1)$,
$(Y_2,\chi_2)$ and $\phi$.

We now introduce the following fermionic symmetry 
(BRST transformations)
\be
\begin{array}{lcl}
Qx^i = i\psi^i, && Q \psi^i = 0,\\
Qy^a = i\tilde{\chi}^a, && Q \tilde{\chi}^a = 0,\\
Q\theta = i\chi_3, && Q \chi_3 = -\phi, \\
QY_{1} =\phi\chi_2, && Q \chi_{1} =  Y_{1}, \\
QY_{2} =-\phi\chi_1, && Q \chi_{2} =  Y_{2}, \\
Q \phi = 0, && \\
\end{array}
\ee
or
\be
\begin{array}{lcl}
Qx^\mu = i\psi^\mu, && Q \psi^\mu = -\phi V^\mu,\\
QY_{1} =\phi\chi_2, && Q \chi_{1} =  Y_{1}, \\
QY_{2} =-\phi\chi_1, && Q \chi_{2} =  Y_{2}, \\
Q \phi = 0, && \\
\end{array}
\ee
in the ambient supermanifold.
Then the action is invariant under this symmetry. Indeed if we write
\be
S=S_1+ S_2,
\ee
where
\bea
S_1 &=& i\phi\mu_3 + \frac{1}{2}\hat{\Omega}_{\mu\nu}\psi^\mu\psi^\nu,\\
S_2 &=&  i Y_1 \mu_1 +  i Y_2 \mu_2  + \sum_{a=1}^3 \left\{
\chi^1  \frac{\del \mu_1}{\del y^a} \tilde{\chi}^a+  \chi^2 \frac{\del \mu_2}{\del y^a}\tilde{\chi}^a
\right\},
\eea
we can show that $S_1$ is BRST closed, namely $QS_1=0$, and
$S_2$ can be written simply as a BRST exact form
\be
S_2 = Q\left\{
i\chi_1\mu_1 + i\chi_2\mu_2
\right\},
\ee
if we use $\frac{\del \mu_{1,2}}{\del x^i} = 0$, etc.
Thus the action is invariant under the BRST symmetry.
Because of the BRST symmetry, the integral 
(\ref{Volume of HK}) is localized at the fixed points 
by using similar arguments to the \Kahler quotient case.

In this section, we discussed how to evaluate the volume 
of the \Kahler and \HK quotient spaces which are 
spanned by spatial coordinates. 
We found that the volume integral is represented 
by an 
integral 
(\ref{modified integral of Kahler}) 
or (\ref{Volume of HK}) over the 
bosonic and fermionic coordinates of the ambient space. 
In the following, we will treat the moduli space of 
BPS vortices as a quotient space defined by moment maps 
imposing BPS constraints on fields instead of spatial 
coordinates. 
Therefore we need to extend the integral over spatial 
coordinates to a path integral over the fields. 
Apart from 
replacing the coordinates by the fields, 
many features of the above arguments in the coordinate 
space will still be valid in evaluating the volume of the 
moduli space of BPS vortices.

\section{Abelian Vortex}

\subsection{Localization of path integral in Abelian case}

In this section, we consider the volume of the moduli 
space of BPS vortices in Abelian ($G=U(1)$) gauge theory 
with $N_f$ Higgs fields on a compact Riemann surface 
$\Sigma_h$ of genus $h$.
Introducing complex coordinates $z=x^1+ix^2$ and 
$\zb=x^1-ix^2$ on $\Sigma_h$, 
the conformally flat metric of the Riemann surface is 
defined as 
\be
ds^2 = g_{z\zb} dzd\zb.
\ee
We also define the \Kahler 2-form 
$\omega=\frac{i}{2} g_{z\zb}dz\wedge d\zb$ from the metric. 
Then the area of the Riemann surface 
$\Sigma_h$ is given by 
\be
{\cal A}=\int_{\Sigma_h}\omega.
\ee

There exists a $U(1)$ gauge field on the Riemann surface 
$\Sigma_h$. 
The field strength is defined in terms of the complex 
coordinates by 
\be
F_{z\zb} = i[D_z,D_\zb],
\ee
where $D_z \equiv \del_z-iA_z$ and 
$D_\zb \equiv \del_\zb-iA_\zb$ 
are covariant derivatives 
for fields with unit $U(1)$ charge 
and 
$A_z$ and $A_\zb$ are 
gauge fields. 

We now define the BPS equations of the Abelian vortex 
\bea
&&F - \frac{g^2}{2}(c-HH^\dag)\omega=0,
\label{AV1}\\
&&D_{\zb} H = D_z H^\dag = 0,
\label{AV2}
\eea
where $g$ is the $U(1)$ gauge coupling, $c$ is a 
 Fayet-Iliopoulos (FI) parameter. 
The Higgs field $H(z,\zb)$ has unit charge and is 
represented by an $N_f$ component vector. 
The covariant derivatives of the Higgs fields are defined by 
$D_{\zb} H\equiv \del_\zb H -iA_\zb H$ and 
$D_{z} H^\dag\equiv \del_z H^\dag +iA_z H^\dag$.
The topological charge (vorticity) $k$ is given by an integral 
of the 2-form field strength $F\equiv F_{z\zb}dz \wedge d\zb$ 
\be
k = \frac{1}{2\pi}\int_{\Sigma_h}F.
\ee

The BPS equations (\ref{AV1}) and (\ref{AV2}) define 
three moment maps 
\bea
&&\mu_r \equiv F - \frac{g^2}{2}(c-HH^\dag)\omega,\\
&&\mu_\zb \equiv D_{\zb} H,\\
&&\mu_z \equiv  D_z H^\dag.
\eea
Using these moment maps, the moduli space ${\cal M}_k$ 
of the vortex with the vorticity $k$ is given by a 
K\"ahler quotient space
\be
{\cal M}_k = \frac{\mu_r^{-1}(0) \cap \mu_z^{-1}(0) 
\cap \mu_\zb^{-1}(0)}{U(1)},
\label{hyper Kahler quotient}
\ee
where $\mu_r^{-1}(0)$ stands for the space of solutions 
which satisfy $\mu_r =0$ and $\frac{1}{2\pi}\int_{\Sigma_h}F=k$, 
etc. 
Precisely speaking, this quotient space is \Kahler but 
not \HK  since the three moment maps do not form triplets 
of the $SU(2)$ isometry. 
However the structure of the three moment maps is very 
similar to the \HK case. So we can utilize the discussions 
of the \HK case in the previous section.

We now introduce fermionic fields $\lambda$, $\psi$ 
to define BRST transformations for fields as follows
\be
\begin{array}{lcl}
QA = i\lambda, && Q \lambda = -d \Phi,\\
Q H = i\psi, && Q\psi = \Phi H,\\
Q H^\dag = -i\psi^\dag, && Q\psi^\dag = \Phi H^\dag,\\
QY = \Phi *\chi, && Q \chi = Y,\\
Q \Phi = 0, && \\
\end{array}
\ee
where we have used form notations  
for the gauge field 
$A \equiv A_z dz+A_\zb d\zb$, 
for the bosonic one-form $Y=Y_z dz+Y_\zb d\zb$ and 
fermionic one-form $\chi=\chi_z dz+\chi_\zb d\zb$ ($*\chi=i(\chi_z dz-\chi_\zb d\zb)$).
The BRST pair of these auxiliary fields $Y$ and $\chi$ 
are $N_f$ component vectors similarly to the Higgs field 
$H$, and will be used as Lagrange 
multipliers of the moment map constraints.
These BRST transformations are nilpotent up to gauge transformations,
namely
$Q^2= -i\delta_{\Phi}$,
where $\delta_{\Phi}$ is the generator of the gauge transformation
with infinitesimal parameter $\Phi$.
Thus if we consider gauge invariant operators only, the BRST transformation $Q$
forms a cohomology for those operators, which is called the ``equivariant cohomology''.
The equivariant cohomology clarifies topological aspects of (topological) field theory considering,
and will play an essential but indirect role of the ``localization'' in the evaluation of the volume.

$\Phi$ is BRST closed itself, so any function of $\Phi$
\be
{\cal O}_0 \equiv {\cal W}(\Phi)
\ee
is also BRST closed. In the sense of the BRST symmetry, 
the 0-form operator becomes a good (physical) observable.
The 0-form observable satisfies the so-called descent 
relation 
\be
d{\cal O}_0 + Q {\cal O}_1=0,
\ee
where
\be
{\cal O}_1 \equiv \frac{\del {\cal W}(\Phi)}{\del \Phi}\lambda.
\ee
This fact means that the integral of ${\cal O}_1$ along a 
closed circle $\gamma$ on $\Sigma_h$ 
\be
\int_\gamma {\cal O}_1
\ee
is BRST closed and a good cohomological observable. 
Similarly, ${\cal O}_1$ satisfies
\be
d{\cal O}_1 + Q {\cal O}_2=0,
\label{eq:descent2}
\ee
where
\be
{\cal O}_2 \equiv i\frac{\del {\cal W}(\Phi)}{\del \Phi}F
+\frac{1}{2}\frac{\del^2 {\cal W}(\Phi)}{\del \Phi^2}
\lambda \wedge \lambda.
\label{eq:operator_descent2}
\ee
Thus the operator 
\be
\int_{\Sigma_h}{\cal O}_2 
\ee
also becomes BRST closed. 
If we choose ${\cal W}(\Phi)=\frac{1}{2}\Phi^2$, we find 
that the integral 
\be
\int_{\Sigma_h}\left[
i\Phi F+\frac{1}{2}\lambda\wedge \lambda
\right]
\ee
is BRST closed. Furthermore, we can see an integral 
\be
\int_{\Sigma_h}\left[
i\Phi H H^\dag\omega +\psi^\dag\psi\omega
\right]
\ee
is also BRST closed, since the integrand can be written 
by the BRST exact form $Q\left[-iH\psi^\dag\omega\right]$.

We would like to calculate the volume of the moduli 
space ${\cal M}_k$, that is the volume of the solutions, 
of the BPS equations (\ref{AV1}) and (\ref{AV2}). 
The volume of the moduli space of the BPS equations 
is obtained from 
the following integral,\footnote{
The normalization of the functional measure 
usually has an ambiguity, but the BRST symmetry 
(supersymmetry) can remove most of the ambiguity. 
} 
although it 
looks like a ``partition function'' 
\be
{\cal V}_k= \int \D \Phi \D^2 Y \D^2 \chi \D^2 A \D^2 
\lambda\D^2 H \D^2 \psi\, 
e^{-S}, 
\label{partition function of BPS eqs}
\ee
where the path integral is taken to satisfy the 
constraint $\frac{1}{2\pi}\int F=k$ 
and the normalization factor $\frac{1}{2\pi}$ comes 
from the volume of the unitary group $U(1)$. 
In the spirit of the previous section, the ``action'' 
$S$ is defined by 
\be
S=S_0+S_1,
\label{eq:action_volume}
\ee
where
\bea
S_0&=& \int_{\Sigma_h}\left[
 i\Phi \mu_r + \frac{1}{2}\lambda\wedge\lambda
+\frac{g^2}{2}\psi^\dag\psi\omega
 \right],\\
S_1&=& t_1 Q\int_{\Sigma_h}
i\chi \wedge * \mu_c,
\label{BRST exact action}
\eea
where $\mu_c \equiv \mu_z dz + \mu_\zb d\zb$.
We can show the action is invariant under the BRST 
symmetry $QS=0$. 
We expect that this path integral defines the 
volume of the moduli space $\M_k$.

If the action includes BRST exact terms with couplings, 
the path integral does not depend on the couplings. 
To see this let us consider the following deformation 
of the integral 
\be
{\cal V}_k'= \int \D \Phi \D^2 Y \D^2 \chi \D^2 A 
\D^2 \lambda \D^2 H \D^2 \psi\,
e^{-S-tQ\Xi}.
\ee
Differentiating it 
with respect to the coupling $t$, we find 
\bea
\frac{\del {\cal V}_k'}{\del t} &=& 
\int \D \Phi \D^2 Y \D^2 \chi  
\D^2 A \D^2 \lambda\D^2 H \D^2 \psi\, (-Q\Xi)
e^{-S-tQ\Xi}\\
&=&-\int \D \Phi \D^2 Y \D^2 \chi  
\D^2 A \D^2 H \D^2 \lambda\D^2 \psi\, Q\left(\Xi 
e^{-S-tQ\Xi}\right)=0,
\eea
where we have used $QS=0$ and invariance of the 
measure under the BRST symmetry.
So the 
integral 
${\cal V}_k$ is independent of the coupling $t_1$ 
in (\ref{BRST exact action}).

Using this coupling independence of the  
integral, we can add the following BRST exact 
term to the action $S$ in Eq.~(\ref{eq:action_volume}) 
without changing the value of the integral 
\be
S_2 = t_2 Q\int_{\Sigma_h} i\chi \wedge * Y.
\ee
By exploiting the coupling independence, we can go to a 
parameter region where the integral can be easily performed: 
let us take the limit $t_1\to 0$ and $t_2\to 1$ of the 
BRST exact terms. Then 
the action becomes 
\bea
S' &=& \int_{\Sigma_h}\left[
 i\Phi \mu_r + \frac{1}{2}\lambda\wedge\lambda
+\frac{g^2}{2}\psi^\dag\psi\omega
 \right]
 + Q\int_{\Sigma_h} i\chi \wedge * Y\\
 &=&  \int_{\Sigma_h}\bigg[
  i\Phi \left\{F-\frac{g^2}{2}(c-HH^\dag)\omega\right\} 
+ \frac{1}{2}\lambda\wedge\lambda 
+\frac{g^2}{2}\psi^\dag\psi\omega\nn\\
  && \qquad\qquad+iY \wedge * Y + i\Phi \chi \wedge \chi
 \bigg].
\label{eq:finalBRSTaction}
\eea
Thus we can use the integral 
\be
{\cal V}_k=\int \D \Phi \D^2 Y  \D^2 \chi \D^2 A 
\D^2 \lambda\D^2 H \D^2 \psi\,
e^{-S'},
\ee
to evaluate the volume of moduli space $\M_k$.

First of all, we wish to integrate out the 
matter fields $H$, $\psi$, $Y$ and $\chi$, whose 
integrals are Gaussian. 
Neglecting the possible anomalies 
coming from the fermionic zero modes of matter fields, 
we obtain 
\be
{\cal V}_k= \int \D \Phi \D^2 A \D^2 \lambda \,
(i\Phi)^{N_f(\dim  \Omega^1\otimes {\cal L}_k 
- \dim  \Omega^0\otimes {\cal L}_k)}
e^{-\int_{\Sigma_h}\bigl[i\Phi(F-\frac{g^2c}{2}\omega)
+ \frac{1}{2}\lambda\wedge\lambda \bigr]},
\ee
where $\dim \Omega^n\otimes {\cal L}_k$ ($n=0,1$) stands 
for the number of holomorphic $n$-forms coupled with 
$U(1)$ gauge field (holomorphic line bundle) of the 
topological charge $k$. 
The Hirzebruch-Riemann-Roch theorem says 
(see e.g.Ref.~\citen{Nakahara:1990th}) 
\be
\dim  \Omega^0\otimes {\cal L}_k 
- \dim  \Omega^1\otimes {\cal L}_k 
= 1-h + \frac{1}{2\pi}\int_{\Sigma_h} F=1-h+k.
\ee
Thus we have 
\be
{\cal V}_k= \int \D \Phi \D^2 A \D^2 \lambda \,
\frac{1}{(i\Phi)^{N_f(1-h+k)}}
e^{-\int_{\Sigma_h}
\bigl[i\Phi(F-\frac{g^2c}{2}\omega)
+ \frac{1}{2}\lambda\wedge\lambda \bigr]}.
\label{eq:gauge_field_integral_volume}
\ee
By using 
$2(1-h)=\frac{1}{4\pi}\int_{\Sigma_h}R^{(2)}$ 
and $ k = \frac{1}{2\pi}\int_{\Sigma_h} F$ 
in terms of 
the curvature 2-form $R^{(2)}$ of the Riemann surface 
and the field strength 2-form $F$, 
we can exponentiate powers of $\Phi$ in 
Eq.~(\ref{eq:gauge_field_integral_volume}) to obtain 
\be
{\cal V}_k= \int \D \Phi \D^2 A \D^2 \lambda \,
e^{-S_{\rm eff}}
.
\label{eq:gauge_field_integral_volume2}
\ee
\be
S_{\rm eff} = S_R + S_F + S_V,
\ee
where
\begin{eqnarray}
S_R &=& \frac{1}{8\pi}\int_{\Sigma_h} 
\log (i\Phi) R^{(2)},\\
S_F &=& 
\int_{\Sigma_h} 
\left[
i\left( 
\Phi +\frac{1}{2\pi i} \log i\Phi \right)F 
+ \frac{1}{2}\lambda\wedge\lambda \right], 
\\
S_V &=&  -i\frac{g^2c}{2}\int_{\Sigma_h}
\Phi\omega
.
\end{eqnarray}
However $S_F$ is not invariant under the BRST symmetry 
(not BRST closed).
Since any regularization scheme should preserve the 
BRST symmetry, this means that we have  overlooked 
contributions from the fermionic zero modes in the 
integrals of fields $\psi, \chi$. 
To recover the contributions from the fermionic zero 
modes, we notice that the BRST closed action must 
take the form (\ref{eq:operator_descent2}) 
given by the descent relation (\ref{eq:descent2}) 
\be
S_{F}' = \int_{\Sigma_h} \left[
i\frac{\del \W_{\rm eff}}{\del\Phi}F+\frac{1}{2}
\frac{\del^2 \W_{\rm eff}}{\del\Phi^2}
\lambda\wedge \lambda
\right],
\label{eq:additional_fermion}
\ee
where 
the tree level term $\frac{1}{2}\Phi^2$ is 
accompanied by the quantum correction 
$\frac{N_f}{2\pi i}\Phi(\log i\Phi -1)$ 
\be
\W_{\rm eff}(\Phi) = \frac{1}{2}\Phi^2 
+ \frac{N_f}{2\pi i}\Phi(\log i\Phi -1).
\ee 
Then we obtain 
\be
S_{F}' = 
\int_{\Sigma_h} 
\left[i\left( 
\Phi +\frac{1}{2\pi i} \log i\Phi \right)F 
+ \frac{\mu(\Phi)}{2}\lambda\wedge\lambda \right], 
\label{eq:corrected_fermion_action}
\ee
where 
\be
\mu(\Phi) = \frac{\del^2 \W_{\rm eff}}{\del\Phi^2} 
= 1+\frac{N_f}{2\pi i \Phi}.
\label{eq:mu}
\ee
The only correction due to (previously neglected) anomalies 
of the fermionic zero modes is changing the coefficient 
of $\frac{1}{2}\lambda\wedge\lambda$ from unity to $\mu(\Phi)$, 
which assures the BRST symmetry of the effective 
action. 

To perform the integration over $A$, 
we decompose the $U(1)$ 
gauge fields into classical configuration and fluctuations.
In terms of the field strength, this means
\be
F = F_{cl} + F_q,
\ee
where $F_{cl}= \frac{2\pi k}{{\cal A}} \omega$, 
which satisfies $\frac{1}{2\pi}\int F_{cl} =k$, 
and $F_q=dA_q$ are quantum fluctuations. 
Integrating the fluctuations $A_q$, we find a constraint 
$d\Phi=0$ on $\Sigma_h$ with Jacobians
from the effective action $\int i\frac{\del \W_{\rm eff}}{\del\Phi}F$.
This means that $\Phi(z,\zb)$ takes a constant value 
$\phi$ everywhere on $\Sigma_h$.

And also, we decompose $\lambda$ into $\lambda=\lambda_0 + \delta\lambda$,
where $\lambda_0$ is a harmonic 1-form and $\delta\lambda$ is fluctuation orthogonal to $\lambda_0$.
Determinants from integrals of $\delta\lambda$ are completely canceled with the Jacobians
induced from the integral of $A_q$. Then, integrals of fermionic zero modes $\lambda_0$ as the harmonic 1-form,
which has $h$ independent components
on $\Sigma_h$, induce a residual contribution $\mu(\phi)^h$ to the integral over the constant mode $\phi$. 

Combining these integral pieces and using $\int_{\Sigma_h}\omega ={\cal A}$ and 
$\int_{\Sigma_h} F = 2\pi k$, 
we finally reduce the path integral into the 
following one-dimensional integral over 
ordinary real number $\phi$ (the 
constant mode of the field $\Phi$) 
\be
{\cal V}_k = \int_{-\infty}^{\infty} \frac{d\phi}{2\pi} 
\frac{\mu(\phi)^h}{(i\phi)^{N_f(1-h+k)}}e^{
i \phi \left( \frac{g^2c}{2}{\cal A} -2\pi k\right)}
 = \int_{-\infty}^{\infty} 
\frac{d\phi}{2\pi} 
\frac{\mu(\phi)^h}{(i\phi)^{N_f(k+1-h)}}
e^{2\pi i \phi B},
\label{eq:zero_mode_integral1}
\ee
where $\mu(\phi)$ is defined in Eq.~(\ref{eq:mu}) and 
\be
B\equiv \frac{g^2c}{4\pi}{\cal A}-k. 
\ee

Now let us evaluate the above integral. 
Since the integrand has a pole at $\phi=0$, we need to 
look for the correct integration contour to avoid the pole. 
The term $\int i\Phi HH^\dagger \omega$ in the action 
 (\ref{eq:finalBRSTaction}) of the path integral reveals 
that we need to choose the contour below the real axis 
in order to assure the convergence of path integral 
of matter fields $H$. 
Namely we should avoid the pole at $\phi=0$ 
counter-clock-wise below the pole. 
Expanding the integrand in powers 
of $\phi$, we can integrate 
term by term
\bea
{\cal V}_k &=& \int_{-\infty-i\e}^{\infty-i\e} 
\frac{d\phi}{2\pi}
 \frac{\left(1+\frac{N_f}{2\pi}
\frac{1}{i\phi}\right)^h}{(i\phi)^{N_f(k+1-h)}}
e^{2\pi i \phi B}\nn\\
&=& \sum_{j=0}^h \binom{h}{j} 
\int_{-\infty-i\e}^{\infty-i\e} \frac{d\phi}{2\pi} 
\left(\frac{N_f}{2\pi}\right)^{h-j}
\frac{1}{(i\phi)^{N_f(k+1-h)+h-j}}
e^{2 \pi i \phi B}\nn\\
&=&
\begin{cases}
{\displaystyle
\sum_{j=0}^h \frac{h!}{j!(h-j)!}
\left(\frac{N_f}{2\pi}\right)^{h-j}
\frac{(2\pi B)^{d-j}}{(d-j)!}}, 
& \text{for } B\geq 0 \text{ and } d\geq h  \\
{\displaystyle
\sum_{j=0}^d  \frac{h!}{j!(h-j)!}
\left(\frac{N_f}{2\pi}\right)^{h-j}
\frac{(2\pi B)^{d-j}}{(d-j)!}}, 
& \text{for } B\geq 0 \text{ and } 0 \le d<h\\
0, & B<0 \text{ or } d<0
\end{cases}
\label{Abelian Nf}
\eea
where we have defined 
$d\equiv kN_f+ (1-h)(N_f-1)$ and used the residue 
integral formula
\be
\int_{-\infty-i\e}^{\infty-i\e}
\frac{d\phi}{2\pi} \frac{1}{(i\phi)^{n+1}} e^{2\pi i \phi B} = 
\begin{cases}
{\displaystyle
\frac{(2\pi B)^n}{n!}}, &
B \ge 0 \text{ and }n\in \Z_{\ge 0}\\
\quad 0, & 
B<0 \text{ or } n\in \Z_{< 0}
\end{cases}.
\label{residue integral}
\ee 
Eq.~(\ref{Abelian Nf}) should give the volume of the 
moduli space of the Abelian vortex with $N_f$ Higgs fields. 
If $B < 0$, namely ${\cal A} < \frac{4\pi}{g^2 c} k$, 
the integral vanishes. 
This means that there is no solution 
for ${\cal A} < \frac{4\pi}{g^2 c} k$. 
This result is in agreement with the bound found in the 
case of ANO vortices ($N_c=N_f=1$), 
which is known as the Bradlow bound.\cite{Bradlow:1990ir}
More interestingly, 
the non-vanishing volume exists 
only if 
$d\geq 0$, namely $k \geq (h-1)\frac{N_f-1}{N_f}$.
So we can choose any 
non-negative 
$k$ for $h=0,1$ 
or $N_f=1$, but $k$ must be sufficiently large 
for $h>1$ in the case of $N_f>1$. 

In this way, the evaluation of the volume of the vortex 
moduli space using the path integral 
reduces finally to the contour integral whose value 
is determined by the residue at the poles. 
The positions of the poles correspond to the fixed points 
of the BRST symmetry. 
This fact reflects the localization theorem in the 
topological field theory path integral with the BRST symmetry.

\subsection{ANO vortices ($N_f=N_c=1$) }

Let us consider the simplest case $N_f=1$, namely the 
ANO vortices.  
In this case, 
Eq.~(\ref{Abelian Nf}) reduces to
\be
{\cal V}_k =
\begin{cases}
{\displaystyle
(2\pi)^{k-h}
\sum_{j=0}^h \frac{h!}{j!(h-j)!}\frac{B^{k-j}}{(k-j)!}}, 
& \text{for } B\geq 0 \text{ and } k\geq h  \\
{\displaystyle
(2\pi)^{k-h}
\sum_{j=0}^k \frac{h!}{j!(h-j)!}\frac{B^{k-j}}{(k-j)!}}, 
& \text{for } B\geq 0 \text{ and } k<h\\
\quad \quad 0, & \text{for } B<0 
\end{cases}
\ee

In our field theoretical evaluation of the volume, 
the integral of $k=0$ sector, that is the sector of 
the flat connections represents the volume of the vacuum 
moduli space. 
Setting $k=0$, we obtain 
\be
{\cal V}_0 = \frac{1}{(2\pi)^h}.
\ee
The net contribution from the $k$ vortex sector is 
obtained by modding out the contribution from the vacuum. 
So if we define $\tilde{{\cal V}}_k \equiv 
{\cal V}_k/{\cal V}_0$, we find for 
${\cal A}\geq \frac{4\pi}{g^2c}k$ 
\be
\tilde{{\cal V}}_k  = (2\pi)^k\sum_{j=0}^{{\rm min}(h,k)} 
\frac{ h!}{j!(k-j)!(h-j)!}
 \left(\frac{g^2c}{4\pi}{\cal A}-  k \right)^{k-j},
\label{eq:ANOvortexModuliVolume}
\ee 
where min$(h,k)$ denotes the smaller one of $h$ or $k$. 
We can compare our result to the previous result of the 
volume of the moduli space 
obtained from the moduli space metric by 
Manton and Nasir\cite{Manton:1998kq}
\be
\Vol({\M_k})  =\pi^k \sum_{j=0}^{{\rm min}(h,k)} 
\frac{(4\pi)^j \left(g^2c{\cal A}- 4\pi k \right)^{k-j} h!}
{j!(k-j)!(h-j)!}. 
\ee 
We find that they are related by just a normalization factor 
\be
\tilde{{\cal V}}_k
= 
\frac{\Vol(\M_k)}{(2\pi)^{k}} .
\ee 
This overall normalization factor can be attributed to 
an arbitrary normalization scale in defining the moduli 
space metric. 
Therefore we find an exact agreement with the 
previous direct calculation.


Finally, we give a comment on asymptotic behavior in the 
large area limit ${\cal A}\to \infty$. 
Let us define the dimensionless area $\tilde {\cal A}$ 
in unit of the intrinsic area of single vortex $4\pi/(g^2c)$ 
\be
\tilde {\cal A} = \frac{g^2c {\cal A}}{4\pi}. 
\label{eq:dimensionless_area}
\ee
In the large area limit , 
the volume of the moduli space behaves 
\be
{\cal V}_k \sim  \frac{({2\pi \A})^k}{k!},
\label{eq:asymptoticVolume}
\ee
which means that the moduli space of $k$ vortices can be 
regarded as the moduli space of well-separated and 
undistinguishable points on $\Sigma_h$ if the area is 
sufficiently large compared to the intrinsic area 
$4\pi/(g^2c)$ of each vortex. 
In other words, the moduli space of $k$ vortices 
${\cal M}_k$ is the symmetric product space of the 
single vortex moduli space ${\cal M}_1$, 
$({\cal M}_1)^k/S_k$ in that large area limit.

\subsection{Abelian semi-local vortices ($N_f> N_c=1$) }

Let us now list the volume of Abelian semi-local vortices 
for $N_f>N_c=1$, in the case of the smaller genus : 
$h=0$ 
(sphere $S^2$) or $h=1$ (torus $T^2$). 
On the sphere $\Sigma_0 =S^2$, the integral becomes 
\be
{\cal V}_k(S^2) =
\frac{(2\pi B)^{kN_f+N_f-1}}
{(kN_f+N_f-1)!}.
\label{eq:u1Nfvolume}
\ee
This expression suggests the dimension of 
${\cal M}_k(S^2)$ is $2kN_f+2(N_f-1)$. 
The contribution from the vacuum is obtained by setting 
$k=0$ 
\be
{\cal V}_0(S^2) =
\frac{(2\pi)^{N_f-1}}
{(N_f-1)!}\times \A^{N_f-1}.
\ee 
As we will see in \S5, the moduli space 
of single semi-local vortices ($N_f>N_c=1$) 
have moduli living on the complex 
projective space $\C P^{N_f-1}$, whose squared radius 
becomes as large as the area $\tilde{\cal A}$. 
These moduli are usually called non-normalizable modes, 
since their mode functions become non-normalizable in the 
limit of infinite area $\tilde{\cal A}\to \infty$. 
The factor $\frac{(2\pi)^{N_f-1}}{(N_f-1)!}$ 
precisely corresponds 
to the volume of the complex projective space $\C P^{N_f-1}$
with a unit radius. 
It is interesting to note that 
the volume of the vacuum moduli ${\cal V}_0(S^2)$ 
suggests the presence of ``non-normalizable'' 
vacuum moduli living on the complex projective space 
$\C P^{N_f-1}$ with the radius $\sqrt{\cal A}$ in 
unit of the vortex size 
$g\sqrt{c/(4\pi)}$.

If we normalize the integral by modding out the 
contribution from the vacuum moduli, we obtain the net 
contribution of $k$-vortex sector
\be
\tilde{{\cal V}}_k(S^2) =
\frac{(2\pi)^{kN_f}}
{\prod_{j=1}^{kN_f}(j+N_f-1)}
\times
\frac{(\A - k)^{kN_f+N_f-1}}{\A^{N_f-1}}.
\ee
In the large area limit $\tilde{\cal A}\to \infty$, the 
normalized integral behaves
\be
\tilde{{\cal V}}_k(S^2) \sim \frac{(2\pi)^{kN_f}}
{\prod_{j=1}^{kN_f}(j+N_f-1)}\A^{k N_f}.
\ee
This behavior suggests the complex dimension of the 
moduli space of vortex is $kN_f$, which agrees with 
the analysis from the index theorem in 
Ref.~\citen{Hanany:2003hp}.

On the torus $\Sigma_1=T^2$, we obtain
\be
{\cal V}_k(T^2) =  \frac{N_f}{2\pi} 
\frac{(2\pi)^{k N_f}}
{(kN_f)!}
\A (\A- k)^{kN_f-1}.
\ee
In this case, the contribution from the vacuum moduli is 
an $\tilde{\cal A}$-independent constant 
${\cal V}_0(T^2)=\frac{N_f}{2\pi}$.
So we can find that $\dim_\C {\cal M}_k(T^2)=kN_f$ 
from the large area limit.

Finally we give a comment on phenomena just at the 
saturated point of the Bradlow limit, namely 
$\tilde{\cal A}=k$ 
(the dissolving limit).
If there is a simple pole in the integrand of the 
residue integral (\ref{Abelian Nf}),
the integral does not vanish even 
at the Bradlow limit, 
$B=\A -k=0$. 
We find that 
the leading term in the dissolving limit is 
that with $j=d$ in (\ref{Abelian Nf}) 
 \be
 {\cal V}_k(\Sigma_h) = \frac{h!}{d!(h-d)!}
\left(\frac{N_f}{2\pi}\right)^{h-d} + {\cal O}(B),
 \ee
where $d=kN_f+(1-h)(N_f-1)$, which must satisfy 
$0 \leq d \leq h$, that is 
$(h-1)-\frac{h-1}{N_f} \leq k \leq (h-1)+\frac{1}{N_f}$.
For the case of the ANO vortex ($N_f=1$), 
the result agrees with the discussions in Ref.~\citen{Manton:2010sa}.


\section{Non-Abelian Vortex}

\subsection{Localization of path integral in non-Abelian case}

In this section, we extend the previous discussions to 
the non-Abelian gauge group $U(N_c)$ case.
The Higgs fields $H$ is now  charged as a fundamental 
representation  of $N_f$ flavors, that is, $H$ is 
represented as an $N_c \times N_f$ matrix.

Three moment maps of the BPS equations are
\bea
\mu_r &=& F - \frac{g^2}{2}(c-HH^\dag)\omega,
\label{eq:unBPS1}\\
\mu_\zb &=& \D_\zb H,
\label{eq:unBPS2}\\
\mu_z &=& \D_z H^\dag,
\label{eq:unBPS3}
\eea
where $\omega$ is the K\"ahler 2-form on $\Sigma_h$.
The moduli space of this non-Abelian vortices is defined 
by the following \HK like quotient space
\be
{\cal M}_k = \frac{\mu_r^{-1}(0) \cap \mu_z^{-1}(0) 
\cap \mu_\zb^{-1}(0)}{U(N_c)},
\ee
where $\mu_r^{-1}(0)$ stands for a space of solutions 
to $\mu_r=0$ with a constraint 
$\frac{1}{2\pi}\int \Tr F=k$, etc.

The BRST transformations are extended as follows: 
\be
\begin{array}{lcl}
QA = i\lambda, && Q \lambda = -d_A \Phi,\\
Q H = i\psi, && Q\psi = \Phi H,\\
Q H^\dag = -i\psi^\dag, && Q\psi^\dag = H^\dag \Phi,\\
QY_z = i\Phi\chi_z && Q \chi_z = Y_z,\\
QY_\zb = -i\chi_\zb\Phi, && Q \chi_\zb = Y_\zb,\\
Q \Phi = 0,&& \\
\end{array}
\ee
where $d_A\Phi\equiv d\Phi-i[A,\Phi]$.
These BRST transformations also are nilpotent up to 
gauge transformations and they yield
the equivariant cohomology.

The cohomological action for the constraint of $\mu_r=0$ 
is given by 
\be
S_0 = \int_{\Sigma_h} \Tr \left[
i\Phi \left\{F-\frac{g^2}{2}(c-HH^\dag)\omega\right\}
+\frac{1}{2}\lambda \wedge \lambda 
+ \frac{g^2}{2} \psi^\dag\psi\omega
\right].
\ee
The action for the other constraints is given by 
\be
S_1 = Q\int_{\Sigma_h} d^2 z \Tr  \left[
\frac{1}{2}g^{z\zb}(\chi_z \mu_\zb + \mu_z \chi_\zb)
\right],
\ee
but we can replace it with the following Gaussian 
action without changing the value of the path integral 
\bea
S_2 &=& Q\int_{\Sigma_h} d^2 z  \Tr  \left[
\frac{1}{2}g^{z\zb}(\chi_z Y_\zb + Y_z \chi_\zb)
\right]\\
&=& \int_{\Sigma_h} d^2 z  \Tr \left[
g^{z\zb}(Y_z Y_\zb +i\Phi \chi_z \chi_\zb)
\right],
\eea
because of the BRST exactness of $S_1$ and $S_2$. 

Using these actions, the integral, which gives the 
volume of the moduli space of the $k$ non-Abelian 
vortices on $\Sigma_h$, is defined by 
\be
{\cal V}_k^{N_c,N_f}(\Sigma_h) = \int \D \Phi \D^2 A 
\D^2 \lambda \D^2 H \D^2 \psi \D^2 Y \D^2 \chi \, 
e^{-S_0-S_2}
\ee
where $N_c$ and $N_f$ are the number of colors and 
flavors, respectively. 
To integrate out the fields, we choose a gauge which 
diagonalizes $\Phi$ as 
$\Phi=\diag(\phi_1,\phi_2,\ldots,\phi_{N_c})$. 
After integrating out $H$, $\psi$, $Y$, $\chi$ 
and off-diagonal pieces of $A$ and $\lambda$ first, 
the integral reduces to the $U(1)^{N_c}$ gauge theory 
\bea
{\cal V}_k^{N_c,N_f} & =&  
\int \prod_{a=1}^{N_c}  (\D \phi_a \D^2 A_a\D^2 \lambda_a) \,
\frac{\prod_{a\neq b}(i\phi_a-i\phi_b)^{\dim \Omega^0\otimes 
{\cal L}_{k_a} \otimes {\cal L}_{k_b}^{-1} 
- \dim \Omega^1\otimes {\cal L}_{k_a}\otimes 
{\cal L}_{k_b}^{-1}}}
{\prod_{a=1}^{N_c}(i\phi_a)^{N_f(\dim \Omega^0\otimes 
{\cal L}_{k_a} - \dim \Omega^1\otimes {\cal L}_{k_a})}} \nn\\
&& \qquad
\times e^{-\sum_{a=1}^{N_c} \int_{\Sigma_h}\bigl[i\phi_a 
(F^{(a)}-\frac{g^2c}{2}\omega)
+\frac{1}{2} \lambda_a\wedge\lambda_a\bigr] },
\label{eq:volume_path_integral_gauge}
\eea
where the diagonal $a$-th $U(1)$ gauge field, field 
strength and gaugino are denoted as $A_a$, $F^{(a)}$ 
and $\lambda_a$, and $k_a$'s are diagonal $U(1)$ 
topological charges 
$\frac{1}{2\pi}\int F^{(a)} = k_a$, which satisfies 
the constraint of the total topological charge 
$k=\sum_{a=1}^{N_c} k_a$.
The contribution of the numerator 
$
\prod_{a\neq b}(i\phi_a-i\phi_b)^{\dim \Omega^0\otimes 
{\cal L}_{k_a} \otimes {\cal L}_{k_b}^{-1}} 
$
is a power of the Vandermonde determinant and 
comes from the integral of the ghost fields which  
are needed to fix the diagonal gauge of $\Phi$. 


To evaluate the infinite dimensional 
functional determinants, we can use the 
Hirzebruch-Riemann-Roch theorem 
\bea
&&\dim  \Omega^0\otimes {\cal L}_{k_a} 
- \dim  \Omega^1\otimes {\cal L}_{k_a}  = 1-h+k_a,\\
&&\dim  \Omega^0\otimes {\cal L}_{k_a}  \otimes 
{\cal L}_{k_b}^{-1}  
- \dim  \Omega^1\otimes {\cal L}_{k_a} 
\otimes {\cal L}_{k_b}^{-1}  = 1-h+k_a-k_b,
\eea
which reduces the infinite dimensional functional 
determinants to the finite ones. 
Similarly to the Abelian case, we can exponentiate 
 powers of $\phi$ and the Vandermonde determinant in 
Eq.~(\ref{eq:volume_path_integral_gauge}) 
in terms of the curvature 2-form $R^{(2)}$ of the 
Riemann surface and the field strength 2-form $F$, 
and express the volume as a path integral over 
$\phi_a$, $A_a$, and $\lambda_a$ 
with the following Abelian effective action 
\be
S_{\rm eff} = S_R + S_F +S_V,
\ee
where
\bea
S_R &=& \frac{1}{8\pi}\int_{\Sigma_h} 
\left(N_f\sum_{a=1}^{N_c}\log (i\phi_a)\right) R^{(2)},\\
S_F &=&\int_{\Sigma_h} \sum_{a=1}^{N_c} \left[i \left(
\phi_a +\frac{N_f}{2\pi i}\log (i\phi_a)
\right)F^{(a)}
+\frac{1}{2}\lambda_a\wedge \lambda_a\right],\\
S_V &=&  -i\frac{g^2c}{2}\int_{\Sigma_h}
\sum_{a=1}^{N_c}\phi_a\omega.
\eea
To maintain the BRST invariance, we should add 
a fermionic contribution coming from anomaly and obtain 
\be
S_F' = \int_{\Sigma_h} \left[i\sum_{a=1}^{N_c}
\frac{\del \W_{\rm eff}}{\del\phi_a}F^{(a)}
+\frac{1}{2}\sum_{a,b=1}^{N_c}\frac{\del^2 
\W_{\rm eff}}{\del\phi_a\del\phi_b}\lambda_a
\wedge \lambda_b
\right],
\ee
where
\be
\W_{\rm eff}(\phi) =\sum_{a=1}^{N_c}\left\{
 \frac{1}{2}\phi_a^2 + \frac{N_f}{2\pi i} 
\phi_a \left(\log (i\phi_a) -1\right)
\right\}
\ee


Using the same argument as that in the previous section, 
we find constraints of $d\phi_a =0$ on $\Sigma_h$, 
due to the integration of the fluctuations of $A_a$ 
and $\lambda_a$. 
So there remains only an integral over constant modes of 
$\phi_a$. 
Let us denote the constant zero modes by $\phi_a$ 
using the same symbol as the field $\phi_a$ itself. 
We can obtain the volume as finite dimensional 
integrals instead of path integrals of fields 
\bea
{\cal V}_k^{N_c,N_f} &\!\!=&\!\!\!\! \sum_{\sum_a k_a =k}  
\int \prod_{a=1}^{N_c} \frac{d\phi_a}{2\pi}
\frac{\mu(\phi)^h\prod_{a \neq b}
(i\phi_a-i\phi_b)^{1-h+k_a-k_b}}
{\prod_{a=1}^{N_c}(i\phi_a)^{N_f(1-h+k_a)}}
e^{2\pi i \sum_{a=1}^{N_c}\phi_aB_a}\\
&\!\!\!\!=&\!\!\!\! \sum_{\sum_a k_a = k}\! (-1)^\sigma \!
\int \prod_{a=1}^{N_c} \frac{d\phi_a}{2\pi}
\frac{\mu(\phi)^h\prod_{a < b}(i\phi_a-i\phi_b)^{2-2h}}
{\prod_{a=1}^{N_c}(i\phi_a)^{N_f(1-h+k_a)}}
e^{2 \pi i \sum_{a=1}^{N_c}\phi_aB_a},
\label{eq:nonabelianVolume}
\eea
where
\be
\mu(\phi) = \det\left|\frac{\del^2 
\W_{\rm eff}(\phi)}{\del\phi_a\del\phi_b}\right|
 = \prod_{a=1}^{N_c}\left(1+\frac{1}{2\pi i} 
\frac{N_f}{\phi_a}\right),
\ee
\be
B_a \equiv 
\tilde{\cal A} -k_a, 
\ee
 and $\sigma =\frac{1}{2}N_c(N_c-1)(1-h)
-\sum_{a<b}(k_a-k_b)$.
The path integral for the non-Abelian gauge theory again 
reduces to the finite dimensional 
residue integral similar to the Abelian case. 
The localization associated with the BRST symmetry 
causes this reduction and the positions of the poles 
correspond to the fixed points.

For the non-Abelian case ($N_c>1$), the integral for 
general genus $h$ is more complicated than the Abelian 
case due to a power of the Vandermonde determinant 
$\det_{a,b}[(i\phi_a)^{b-1}]=\prod_{a < b}(i\phi_a-i\phi_b)$. 
So let us concentrate the calculation of the integral 
for the small genus case ($h=0,1$). For $h=0$, we obtain 
\be
{\cal V}_k^{N_c,N_f}(S^2) =  \sum_{\sum_a k_a = k} 
(-1)^\sigma \int \prod_{a=1}^{N_c} \frac{d\phi_a}{2\pi}
\frac{\prod_{a < b}(i\phi_a-i\phi_b)^{2}}
{\prod_{a=1}^{N_c}(i\phi_a)^{N_f(k_a+1)}}
e^{2\pi i \sum_{a=1}^{N_c}\phi_aB_a}.
\ee 
Let us define coefficients $a_{l_1,l_2,\dots,l_{N_c}}$ 
in expansion of 
the square of the Vandermonde determinant 
in powers of $i\phi_a$ 
\be
\prod_{a < b}(i\phi_a-i\phi_b)^{2} 
= \sum_{\sum l_a 
= N_c(N_c-1)} a_{l_1,l_2,\dots,l_{N_c}}
(i\phi_1)^{l_1}(i\phi_2)^{l_2}
\cdots(i\phi_{N_c})^{l_{N_c}}.
\ee
Then we have
\be
{\cal V}_k^{N_c,N_f}(S^2) =  \sum_{\sum_a k_a = k} 
(-1)^\sigma 
\sum_{\sum l_a = N_c(N_c-1)} 
a_{l_1,l_2,\dots,l_{N_c}} 
\prod_{a=1}^{N_c} F_{N_f} (k_a,l_a;{\cal A}),
\label{partition function on sphere}
\ee
where
\bea
F_{N_f} (k_a,l_a;{\cal A}) &=&  \int  \frac{d\phi_a}{2\pi}
\frac{e^{2\pi i \phi_aB_a}}{(i\phi_a)^{N_f(k_a+1)-l_a}}\nn\\
&=&
\frac{
\left(2\pi (\A-  k_a)\right)^{N_f(k_a+1)-l_a-1}}
{(N_f(k_a+1)-l_a-1)!}
.
\eea

The contribution from the vacuum moduli space is obtained 
from the $k=0$ case, namely all $k_a=0$. 
Then we obtain 
\begin{multline}
{\cal V}_0^{N_c,N_f}(S^2) =  (-1)^{N_c(N_c-1)/2}
(2\pi)^{N_c (N_f-N_c)}\\
\times \sum_{\sum l_a = N_c(N_c-1)} 
\frac{a_{l_1,l_2,\dots,l_{N_c}}}{\prod_{a=1}^{N_c} (N_f-l_a-1)!}
\A^{N_c (N_f-N_c)}.
\end{multline}
It it difficult to evaluate the above formula for 
general $N_c$. 
However, 
results for smaller values of $N_c$ suggest the following 
formula 
\be
{\cal V}_0^{N_c,N_f}(S^2) = N_c! \times \Vol(G_{N_c,N_f})
 \A^{N_c (N_f-N_c)},
\ee
where $\Vol(G_{N_c,N_f})$ is the volume of the Grassmannian 
$G_{N_c,N_f}$ with a unit radius. (See Appendix.)
We will check this conjecture for the smaller $N_c$ cases 
in the next section.
Recalling the fact that the complex dimension of the 
Grassmannian is $N_c(N_f-N_c)$, we see 
${\cal V}_0^{N_c,N_f}(S^2)$ represents the contribution 
from the vacuum moduli, including non-normalizable modes, 
in the non-Abelian Yang-Mills-Higgs 
system.\footnote{The factor $N_c!$ comes from the Weyl 
permutation of the Higgs vevs.}

Using this expression (\ref{partition function on sphere}), 
we naively expect the large area behavior as
\begin{multline}
{\cal V}_k^{N_c,N_f}(S^2) \sim 
\sum_{\sum_a k_a = k} 
(-1)^\sigma
\sum_{\sum l_a = N_c(N_c-1)} 
\frac{a_{l_1,l_2,\dots,l_{N_c}}}
{\prod_{a=1}^{N_c} (N_f(k_a+1)-l_a-1)!}\\
\times
(2\pi \A)^{k N_f+N_c(N_f-N_c)}.
\end{multline}
So by modding out the vacuum contribution, 
we find the volume of the $k$-vortex sector diverges as 
$kN_f$-th power of $\A$. 
However, we will see this observation is too naive 
for the case of the local vortex $N_f=N_c$. 
In this case, the higher powers of $\A$ are non-trivially 
cancelled with each other due to a property of 
the coefficients $a_{l_1,l_2,\dots,l_{N_c}}$ in the 
expansion of the Vandermonde determinant.

For the torus case ($h=1$), the integral becomes 
simpler because of the absence of the Vandermonde 
determinant.
We can integrate each $\phi_a$'s independently. 
Thus we obtain
\be
{\cal V}_k^{N_c,N_f}(T^2) =\sum_{\sum_a k_a = k} (-1)^\sigma 
 \prod_{a=1}^{N_c} {\cal V}^{1,N_f}_{k_a}(T^2),
\label{non-Abelian on torus}
\ee
where ${\cal V}^{1,N_f}_{k_a}(T^2)$ is the volume of moduli 
space of the Abelian vortex with $N_f$ flavors,
which has vorticity $k_a$, on the torus
\be
{\cal V}_{k_a}^{1,N_f}(T^2) = \frac{N_f}{2\pi} 
\frac{(2\pi)^{k_a N_f}}
{(k_a N_f)!}
\A (\A - k_a )^{k_a N_f-1}.
\label{Abelian Nf partition function}
\ee
Substituting (\ref{Abelian Nf partition function}) 
into (\ref{non-Abelian on torus}), we find that the 
general formula 
\be
{\cal V}_k^{N_c,N_f}(T^2) = 
(2\pi)^{k N_f}  \left(\frac{N_f}{2\pi} \A \right)^{N_c} 
\sum_{\sum_a k_a = k} (-1)^\sigma
 \prod_{a=1}^{N_c} \frac{(\A-k_a)^{k_a N_f-1}}{(k_a N_f)!}.
\ee
In particular, if we set $N_c=N_f=N$ (local vortex), we obtain
\be
{\cal V}_k^{N,N}(T^2) = 
(2\pi)^{k N}  \left(\frac{N}{2\pi} \A \right)^{N} 
\sum_{\sum_a k_a = k} (-1)^\sigma
 \prod_{a=1}^{N} \frac{(\A-k_a)^{k_a N-1}}{(k_a N)!}.
\ee
And in the large area limit $\A\to\infty$ the volume of the moduli space of the local vortices on the torus
behaves
\be
{\cal V}_k^{N,N}(T^2) \sim
(2\pi)^{(k-1) N}  N^N \A^{k N}
\sum_{\sum_a k_a = k} (-1)^\sigma
 \prod_{a=1}^{N} \frac{1}{(k_a N)!}.
\label{large A on torus}
\ee


In the following subsections, we evaluate the volume of 
moduli space of the non-Abelian vortices 
for some restricted number of $N_c$, $N_f$ and in 
particular, smaller genus cases  of $h=0$ or $h=1$ for 
simplicity, and look in detail at properties of 
the volume.

\subsection{On sphere $S^2$}

We first evaluate the spherical case for $N_c=2$. 
The integral becomes
\be
{\cal V}_k^{2,N_f}(S^2) =  \sum_{k_1+k_2 = k} (-1)^\sigma
\int \frac{d\phi_1}{2\pi} \frac{d\phi_2}{2\pi}
\frac{(i\phi_1-i\phi_2)^{2}}
{(i\phi_1)^{N_f(k_1+1)}(i\phi_2)^{N_f(k_2+1)}}
e^{2\pi i \phi_1B_1+2\pi i\phi_2B_2},
\ee
where $\sigma=k+1$.
The square of the Vandermonde determinant has the 
following expansion: 
\be
(i\phi_1-i\phi_2)^2 = (i\phi_1)^2 - 2(i\phi_1)(i\phi_2) 
+ (i\phi_2)^2.
\ee
Using the formula (\ref{partition function on sphere}), 
we obtain 
\begin{multline}
{\cal V}_k^{2,N_f}(S^2) = \sum_{\sum_a k_a = k} 
(-1)^\sigma (
F_{N_f} (k_1,2;{\cal A})F_{N_f} (k_2,0;{\cal A})\\
 -2F_{N_f} (k_1,1;{\cal A})F_{N_f} (k_2,1;{\cal A})\\
+F_{N_f} (k_1,0;{\cal A})F_{N_f} (k_2,2;{\cal A})
)
\end{multline}

For $k=0,1,2,3$, we obtain 
\begin{align}
{\cal V}_0^{2,N_f}(S^2) &=  \frac{2!}{(N_f-1)!(N_f-2)!}
(2\pi \A)^{2(N_f-2)} \sim {\cal O}(\tilde{\cal A}^{2N_f-4}),\nn\\
{\cal V}_1^{2,N_f}(S^2) &=  
\frac{(2\pi)^{3N_f-4}}{(2N_f-1)(N_f-1)!(2N_f-3)!}
\A^{N_f-3}(\A-1)^{2N_f-3}
\nn\\
&\qquad \times\left((N_f-2)\A^2+2(N_f+1)\A+(N_f-2)\right)
\sim {\cal O}(\tilde{\cal A}^{3N_f-4}),\nn\\
{\cal V}_2^{2,N_f}(S^2) &= 2(2\pi)^{4N_f-4}\Bigg[
\frac{-2}{(N_f-1)!(3N_f-1)!}
\A^{N_f-3}(\A-2)^{3N_f-3}\nn\\
&\qquad \times\left((2N_f^2-2N_f+1)\A^2
+2(2N_f+1)(N_f-1)\A+2(N_f-1)(N_f-2)\right)\nn\\
&\qquad +\frac{1}{(2N_f-1)!(2N_f-2)!}(\A-1)^{4N_f-4}\Bigg]
\sim {\cal O}(\tilde{\cal A}^{4N_f-4}),\nn\\
{\cal V}_3^{2,N_f}(S^2) &= 2(2\pi)^{5N_f-4}\Bigg[
\frac{1}{(4N_f-1)!(N_f-1)!}\A^{N_f-3}(\A-3)^{4N_f-3}\nn\\
&\qquad \times\left((9N_f^2-5N_f+1)\A^2
+6(3N_f+1)(N_f-1)\A+9(N_f-1)(N_f-2)\right)\nn\\
&\qquad + \frac{1}{(3N_f-1)!(2N_f-1)!}
(\A-1)^{2N_f-3}(\A-2)^{3N_f-3}\nn\\
&\qquad \times\left((N_f^2-5N_f+2)\A^2+2(N_f^2+6N_f-3)\A
+N_f^2-13N_f+6\right)
\Bigg]
\nn\\
&\sim {\cal O}(\tilde{\cal A}^{5N_f-4}).
\label{eq:n=2SemiLocalVolume}
\end{align}
where we show the asymptotic powers of $\tilde {\cal A}$. 
Noting that 
$\Vol(G_{2,N_f}) = \frac{(2\pi)^{2(N_f-2)}}{(N_f-1)!(N_f-2)!}$, 
we can write 
\be
{\cal V}_0^{2,N_f}(S^2) = 2! \Vol(G_{2,N_f}) \A^{2(N_f-2)}.
\label{eq:n=2vacuumVolume}
\ee
In general, in the large $\tilde{\cal A}$ limit, the integral 
behaves as $k N_f +2(N_f-2)$-th power of $\tilde{\cal A}$,
in which $2(N_f-2)$-th power of $\tilde{\cal A}$ comes from the 
vacuum contribution ${\cal V}^{2,N_f}_0$.
 From Eq.~(\ref{eq:n=2SemiLocalVolume}), we conjecture 
the asymptotic power of $\tilde{\cal A}$ for general $N_f$ as 
\be
{\cal V}_k^{2,N_f}(S^2) \propto \A^{kN_f+2(N_f-2)}.
\ee
Thus,
in the large area limit $\tilde{\cal A}\to \infty$, the normalized 
integral behaves
\be
\frac{{\cal V}_k^{2,N_f}(S^2)}{{\cal V}_0^{2,N_f}(S^2)} 
\propto
{\tilde{\cal A}^{kN_f}}. 
\ee

If we specialize to the $N_c=N_f$ case, namely $N_f=2$, the 
integral reduces to
\bea
{\cal V}_0^{2,2}(S^2) &=& 2, \nn\\
{\cal V}_1^{2,2}(S^2) &=& 2 \times (2\pi)^2(\A-1),\nn\\
{\cal V}_2^{2,2}(S^2) &=& 2 \times \frac{(2\pi)^4}{2!}
\left(\A^2-\frac{20}{6}\A+\frac{17}{6}\right),\nn\\
{\cal V}_3^{2,2}(S^2) &=& 2 \times \frac{(2\pi)^6}{3!}
\left(\A^3-7\A^2+\frac{331}{20}\A-\frac{793}{60}\right),\nn\\
{\cal V}_4^{2,2}(S^2) &=& 2  \times \frac{(2\pi)^8}{4!}
\left(\A^4-12\A^3+\frac{818}{15}\A^2-\frac{2336}{21}\A
+\frac{18047}{210}\right).\nn\\
\eea
We expect that the integral behaves in general
\be
{\cal V}_k^{2,2}(S^2) = 2\times \frac{(2\pi)^{2k} \A^k}{k!} 
+ {\cal O}(\A^{k-1}).
\label{eq:n=2LocalVortexVolume}
\ee
Thus,
in the large area limit $\tilde{\cal A}\to \infty$, the normalized 
integral behaves
\be
\frac{{\cal V}_k^{2,2}(S^2)}{{\cal V}_0^{2,2}(S^2)} 
\sim \frac{(2\pi)^{2k}\tilde{\cal A}^{k}}{k!}.
\ee

Similarly, we can evaluate for $N_c=3$.
For $k=0,1$, we obtain
\bea
{\cal V}_0^{3,N_f}(S^2) &=& 
\frac{3!2!}{(N_f-1)!(N_f-2)!(N_f-3)!}
(2\pi\A)^{3(N_f-3)}\sim {\cal O}(\tilde {\cal A}^{3(N_f-3)})\nn\\
{\cal V}_1^{3,N_f}(S^2) &=&  
\frac{6(2\pi)^{4N_f-9}}{(2N_f-1)(2N_f-1)!(2N_f-3)!}
\A^{2N_f-8}(\A-1)^{2N_f-5}\nn\\
&&
\times\Big((N_f-2)(N_f-3)\A^4+4(N_f+1)(N_f-3)\A^3
+6(N_f^2-N_f+4)\A^2\nn\\
&&
+4(N_f+1)(N_f-3)\A+(N_f-2)(N_f-3)\Big)
\sim {\cal O}(\tilde{\cal A}^{N_f+3(N_f-3)}). 
\label{eq:n=3SemiLocalVolume}
\eea
Since 
$\Vol(G_{3,N_f})= \frac{(2\pi)^{3(N_f-3)}2!}
{(N_f-1)!(N_f-2)!(N_f-3)!}$, we can write
\be
{\cal V}_0^{3,N_f}(S^2) = 3! \Vol(G_{3,N_f}) \A^{3(N_f-3)}. 
\label{eq:n=3vacuumVolume}
\ee
 Eqs.~(\ref{eq:n=2vacuumVolume}) and (\ref{eq:n=3vacuumVolume}) 
suggest that the integral of the vacuum sector $k=0$ 
for general $N_c$ and $N_f$ is given by 
\be
{\cal V}_0^{N_c,N_f}(S^2) 
= N_c! \Vol(G_{N_c,N_f}) \A^{N_c(N_f-N_c)}. 
\label{eq:nVacuumVolume}
\ee
 Eqs.~(\ref{eq:n=2SemiLocalVolume}) and 
(\ref{eq:n=3SemiLocalVolume}) also suggest 
that the asymptotic power of $\tilde{\cal A}$ for general 
$N_c$ and $N_f$ is given by 
\be
{\cal V}_k^{N_c,N_f}(S^2) \propto \A^{kN_f+N_c(N_f-N_c)}.
\label{eq:volumeGeneralNcNf}
\ee
Thus, 
in the large area limit $\tilde{\cal A}\to \infty$, the normalized 
integral behaves 
\be
\frac{{\cal V}_k^{N_c,N_f}(S^2)}{{\cal V}_0^{N_c,N_f}(S^2)} 
\propto {\tilde{\cal A}^{kN_f}}.
\ee
It is interesting to note that the highest power of 
$\tilde{\cal A}$ is independent of $N_c$.

If we specialize to $N_c=N_f=3$, we obtain 
\bea
{\cal V}_0^{3,3}(S^2) &=& 3! ,\nn\\
{\cal V}_1^{3,3}(S^2) &=& 3!  \times \frac{(2\pi)^3}{2}(\A-1),\nn\\
{\cal V}_2^{3,3}(S^2) &=& 3!  \times \frac{(2\pi)^6}{2^22!}
\left(\A^2-\frac{46}{15}\A+\frac{36}{15}\right),\nn\\
{\cal V}_3^{3,3}(S^2) &=& 3!  \times \frac{(2\pi)^9}{2^33!}
\left(\A^3-\frac{31}{5}\A^2+\frac{3641}{280}\A
-\frac{23249}{2520}\right).
\label{eq:n=3LocalVortexVolume}
\eea
Eqs.~(\ref{eq:n=2LocalVortexVolume}) and 
(\ref{eq:n=3LocalVortexVolume}) suggest that the 
integral of local vortices for general $N_c=N_f=N$ 
case is asymptotically given by 
\be
{\cal V}_k^{N,N}(S^2) 
\sim \frac{N!}{k!}\left(\frac{(2\pi)^N}{N-1}\tilde{\cal A}\right)^k,
\label{eq:nLocalVortexVolume}
\ee
in the large area limit $\tilde{\cal A}\to \infty$.

\subsection{On torus $T^2$}

For $N_c=2$, the integral is
\be
{\cal V}_k^{2,N_f}(T^2) = 
(2\pi)^{k N_f}  \left(\frac{N_f}{2\pi} \A \right)^{2} 
\sum_{\sum_a k_1+k_2 = k} 
  \frac{(\A-k_1)^{k_1 N_f-1}}{(k_1 N_f)!}
\frac{(\A-k_2)^{k_2 N_f-1}}{(k_2 N_f)!},
\ee
where we have ignored the overall sign of the integral. 
Assuming $\A>k$, we obtain concretely
\bea
{\cal V}_0^{2,N_f}(T^2) &=& \left(\frac{N_f}{2\pi}\right)^{2} , \nn\\
{\cal V}_1^{2,N_f}(T^2) &=& \left(\frac{N_f}{2\pi}\right)^{2} 
\times 2(2\pi)^{N_f}\frac{\A(\A-1)^{N_f-1}}{N_f!},\nn\\
{\cal V}_2^{2,N_f}(T^2) &=&  \left(\frac{N_f}{2\pi}\right)^{2} 
\times (2\pi)^{2N_f} \left[
2\frac{\A(\A-2)^{2N_f-1}}{(2N_f)!}
+\left(\frac{\A(\A-1)^{N_f-1}}{N_f!}\right)^2
\right]
,\nn\\
{\cal V}_3^{2,N_f}(T^2) &=& \left(\frac{N_f}{2\pi}\right)^{2}
\times2(2\pi)^{3N_f} \left[
\frac{\A(\A-3)^{3N_f-1}}{(3N_f)!}
+\frac{\A(\A-1)^{N_f-1}}{N_f!}\frac{\A(\A-2)^{2N_f-1}}{(2N_f)!}
\right].\nn\\
\eea 
In the large area limit $\A \to \infty$, 
the integral always behaves as $kN_f$-th power of 
$\A$ for the semi-local and local vortices.

\section{Effective Lagrangian of Vortices}

In this section we study the effective Lagrangian 
of vortices on Riemann surfaces
to define the moduli space metric of vortices, and 
compute the volume of moduli space in order to compare 
with the results of topological field theory. 
We here will consider only sphere topology for the 
Riemann surface and use the strong coupling limit 
$g^2\to\infty$ to obtain the leading behavior of 
the moduli space volume for large area 
${\cal A}\to \infty$. 
To understand the Bradlow limit for small ${\cal A}$, 
we also study the case of finite $g^2$ for a limited 
class of moduli space with single vortex. 

\subsection{Effective Lagrangian and Moduli Space Metric
}

Let us consider a $(2+1)$-dimensional space-time 
with the line element
\begin{eqnarray}
ds^2
= -dt^2+\sigma [(dx)^2+(dy)^2] 
= -dt^2+g_{z \bar z} dz d\bar z 
=g_{\mu\nu}dx^\mu dx^\nu, 
\label{eq:line_element} 
\end{eqnarray}
where the conformal factor and the complex coordinate 
are denoted as $\sigma=g_{z\bar z}$ and $z=x+iy$, 
respectively. 
We will denote the time coordinate $t$ and 
the spacial coordinates $x, y$ 
by $0$ and $i, j=1,2$, respectively, 
and space-time coordinates by 
$\mu, \nu =0, 1, 2$. 
The area ${\cal A}$ of the Riemann surface is given by the 
conformal factor $\sigma$ as 
\begin{eqnarray}
\int dx dy \, \sigma ={\cal A}. 
\label{eq:areaConformalfactor} 
\end{eqnarray}
 For the sphere, $\sigma$ 
can be chosen with $z=x+iy$ 
as 
\begin{eqnarray}
\sigma =\frac{\cal A}{\pi (1+|z|^2)^2}, 
\quad z \in \C. 
\label{eq:confFactorSphere} 
\end{eqnarray}

We are interested in a $U(N_{c})$ gauge theory in 
$(2+1)$-dimensional space-time with gauge fields 
$A_\mu$ as  $N_c \times N_c$ matrices 
and $N_{f}$ Higgs fields 
in the fundamental representation of the $SU(N_{c})_C$ 
$H$ as an $N_c \times N_f$ matrix. 
The Lagrangian of the theory reads 
\begin{eqnarray}
L&=&\int dx dy \sqrt{-{\rm det}(g_{\mu\nu})} \; \mathcal L
=\int dt dx dy  \; \sigma \; \mathcal L
=T-V,
\label{eq:unlagrangian1} 
\\
\mathcal L &=& \; {\rm Tr}\Bigl[
- \frac{1}{2g^2} F_{\mu \nu}F^{\mu \nu} 
+ {\cal D}_\mu H ({\cal D}^\mu H)^\dagger 
-
\frac{g^2}{4} \left( H H^\dagger - c{\bf 1}_{N_c} \right)^2 
\Bigr],
\label{eq:uNLagrangianDensity}
\\
T&=&
\int dx dy \; {\rm Tr}\left[
\sigma D_t H (D_t H)^\dagger 
+
\frac{1}{g^2}(F_{ti})^2
\right],
\label{eq:unkineticTerm} 
\\
V &=& 
\int dx dy \; {\rm Tr}\left[\frac{1}{g^2\sigma}(F_{12})^2
+D_i H (D_i H)^\dagger 
+ 
\frac{g^2\sigma}{4} ( H H^\dagger - c )^2 \right].
\label{eq:unpotentialTerm} 
\end{eqnarray}
Our notation is 
$A_\mu= A_\mu^I t^I$ with the normalization 
\begin{eqnarray}
{\rm Tr} ( t^I t^J ) = \frac{1}{2} \delta^{IJ}.
\end{eqnarray}
As is well known, this Lagrangian Eq.\,\eqref{eq:uNLagrangianDensity} 
can be embedded into a supersymmetric theory with 
eight supercharges.

The Bogomol'nyi bound for the energy $E$ is obtained 
by completing the square 
\begin{eqnarray}
E &=& 
\int dx dy \; {\rm Tr} \left[4D_{\bar z}H D_z H^\dagger 
+\frac{1}{g^2\sigma}
\left(F_{12}-\frac{g^2\sigma}{2} (c - H H^\dagger )\right)^2
+c F_{12}\right]
\nonumber \\
&\ge& c \int dx dy \; {\rm Tr} (F_{12})=2\pi c k, 
\label{eq:bogomolnyiBound}
\end{eqnarray}
where the topological charge $k$ is the 
vorticity, namely the number of vortices. 
The bound is saturated if the BPS equations 
(\ref{eq:unBPS1}), (\ref{eq:unBPS2}) and (\ref{eq:unBPS3}) 
are satisfied. 
For the Abelian gauge theory, we should just choose 
$N_c=1$.


By rewriting the gauge fields in 
terms of a matrix $S\in GL(N_c, \mathbb{C})$ 
\begin{eqnarray}
A_{\bar z} &=& -iS^{-1}\partial_{\bar z}S,
\label{eq:Smatrix}
\end{eqnarray}
we can solve the BPS equations (\ref{eq:unBPS2}) and 
(\ref{eq:unBPS3}) 
for Higgs field $D_{\bar z}H=0$ by
\begin{eqnarray}
H=S^{-1} H_0 (z), 
\label{eq:unmoduliMatrix}
\end{eqnarray}
with the holomorphic $N_c\times N_f$ matrix $H_0$ 
which is called the moduli 
matrix\cite{Eto:2005yh,Eto:2006mz,Eto:2006uw,Eto:2007aw,Eto:2009wq,
Fujimori:2010fk}. 
Also, by defining positive definite Hermitian matrix 
$\Omega \in GL(N_c,  \mathbb{C})$ 
\begin{eqnarray}
\Omega= S S^\dagger,
\label{eq:omega}
\end{eqnarray}
we can rewrite the remaining BPS equation (\ref{eq:unBPS1}) 
as 
\begin{eqnarray}
\partial_{\bar z}(\Omega \partial_{ z} \Omega^{-1})
=\frac{g^2}{4}
\left( H_0 H_0^\dagger\Omega^{-1}-c \mathbf 1_N \right)\sigma, 
\label{eq:masterEqUn}
\end{eqnarray}
which is called the master equation. 
The master equation is expected to give a unique 
solution, once the moduli matrix $H_0$ is given\cite{Eto:2006pg}. 
The vortex number $k$ in Eq.(\ref{eq:bogomolnyiBound}) 
can be rewritten in terms of $\Omega$ by using 
$F_{z\bar z}=iS^{-1}\partial_{\bar z}
(\partial_z \Omega \Omega^{-1})\Omega S^{\dagger -1}$ as 
\begin{eqnarray}
2\pi  k  
&=&-2i  \int dx dy \, \Tr(F_{z\bar z})
=2  \int dx dy \, \partial_{\bar z}\partial_z 
\log\det\Omega 
\nonumber \\
&=&
-i\oint dz \, \partial_z \log\det\Omega,
\label{eq:vorticity}
\end{eqnarray}
where the contour integration is counter-clock-wise 
and $\partial_z \log\det\Omega$ is 
not single valued. 
Eq.(\ref{eq:vorticity}) specifies the boundary condition 
in choosing the moduli matrix $H_0$ and 
in solving the master equation (\ref{eq:masterEqUn}). 
The definitions (\ref{eq:Smatrix}) and (\ref{eq:omega}) 
imply that the master equation is covariant under 
the holomorphic transformation $V(z)$:
\begin{eqnarray}
H_0(z) \to V(z)H_0(z), \quad 
S \to V(z)S, 
\quad V(z)\in GL(N_c, \mathbb{C}).
\label{eq:Vtransformation}
\end{eqnarray}
Therefore there exists a one-to-one correspondence 
between the equivalence classes 
$H_0\sim V H_0$ and points on the moduli space.

General formula for effective Lagrangian 
$L_{\rm eff}$ on vortices 
has been worked out\cite{Eto:2006uw,Fujimori:2010fk}. 
Here we generalize the method to a curved compact 
manifold such as the Riemann surfaces. 
Assuming slow motion of moduli parameters, effective 
Lagrangian on a soliton background can be obtained 
up to the second order in derivative. 
After subtracting the term in the zero-th order in 
derivative (static energy $2\pi c k$ of the soliton multiplied 
by $-1$), 
the effective Lagrangian on the curved manifold 
with the conformal factor $\sigma$ is given by 
\begin{eqnarray}
L_{\rm eff}+2\pi c k &=&
\int_{\Sigma_h} dx dy {\rm Tr}\Bigl[\sigma \; 
\delta_t^\dagger \left\{\Omega^{-1}\delta_t H_0H_0^\dagger\right\}
\nonumber \\
&+&\frac{4}{g^2}\partial_{\bar z}
\left\{\partial_z(\Omega^{-1} \delta_t^\dagger \Omega)
\Omega^{-1}\delta_t\Omega\right\}\Bigr], 
\label{eq:effectiveLagrangian1}
\end{eqnarray}
where the first term comes from the kinetic term 
of the Higgs field $\sigma D_t H (D_t H)^\dagger$, 
and the second term from the 
electric field $\frac{1}{g^2}(F_{ti})^2$. 
The time derivative through (anti-)holomorphic 
moduli $\phi^i$ (${\bar \phi}^i$) is denoted as 
\begin{eqnarray}
\delta_t = \dot \phi^i \frac{\partial}{\partial \phi^i}, 
\quad 
\delta_t^\dagger = \dot{\bar \phi}^i 
\frac{\partial}{\partial \bar \phi^i}.
\label{eq:holomorphicModuliDerivative}
\end{eqnarray}

In the strong coupling limit $g^2\to\infty$, 
the master equation (\ref{eq:masterEqUn}) 
can be solved algebraically 
\begin{eqnarray}
\Omega = \frac{H_0H_0^\dagger}{c}, 
\quad g^2\to\infty
.
\label{eq:masterEqStrongCoupling}
\end{eqnarray}
Therefore the effective Lagrangian (\ref{eq:effectiveLagrangian1}) 
is reduced to 
a simple formula for the K\"ahler potential $K$ 
\begin{eqnarray}
L_{\rm eff}+2\pi c k=\delta_t^\dagger \delta_t K, 
\label{eq:effectiveLagrangianStrongCoupling}
\end{eqnarray}
where
\begin{eqnarray}
K=c\int_{\Sigma_h} dx dy \; \sigma \; 
 \log\det(H_0H_0^\dagger)
.
\label{eq:kahlerPotential}
\end{eqnarray}

\subsection{Moduli Space Metric of Abelian Semi-Local Vortices 
}

Let us first work out the effective Lagrangian 
of Abelian semi-local vortices ($N_f>N_c=1$) 
on the sphere, using the strong 
coupling $g^2\to\infty$. 

 Moduli matrices should be chosen to satisfy 
the boundary condition (\ref{eq:vorticity}). 
If we consider the large enough area ${\cal A}\to \infty$, 
the solution $\Omega$ of the master equation should approach 
$\Omega \to H_0H_0^\dagger/c$ asymptotically $z\to\infty$.  
Therefore the boundary condition (\ref{eq:vorticity}) 
is satisfied by requiring that at least one of components 
of moduli matrix to be a polynomial of order $k$, 
and all other components to be at most of order $k$ 
\begin{eqnarray}
H_0^{(k)}(z) =\sqrt{c}\left(\sum_{j=0}^{k}a^{(1)}_{j}z^j, \cdots, 
\sum_{j=0}^{k}a^{(N_f)}_{j}z^j\right), 
\label{eq:u1moduliMatrix_kvortex} 
\end{eqnarray}
where at least one of the coefficients of the $k$-th power 
is non-vanishing: $a^{(j)}_{k}\not=0$. 
 If we adiabatically deform the base Riemann surface, this should still 
be the appropriate moduli matrix, until possible critical 
value is reached where the solution of the master equation 
ceases to exist. 
We expect that the Bradlow bound\cite{Bradlow:1990ir} 
gives such a critical value of the area. 

Compared to the usual noncompact plane, we emphasize 
two new features of the vortex moduli on the 
compact Riemann surfaces which are realized in the moduli 
matrix (\ref{eq:u1moduliMatrix_kvortex}). 
Firstly we allow the leading power of $z$ to be in any 
components. 
If we use global $SU(N_f)$ rotations combined with the 
$V$-transformations, it is possible to 
place the leading power to be in a particular component, 
say in the first component. 
This form is the usual choice for the moduli matrix on 
noncompact flat plane \cite{Eto:2007yv}. 
These new $N_f-1$ moduli may be regarded as an orientation 
of the vacuum at infinity and are nonnormalizable on 
noncompact plane:  $(a_k^{(1)}, \cdots, a_k^{(N_f)})/a_k^{(1)}$ 
after dividing out by $V$-transformations. 
These $N_f-1$ extra complex moduli 
parameters are present even in the case of vacuum 
($k=0$) on compact Riemann surfaces. 
Secondly the additional $N_f-1$ ``size'' moduli are retained 
on compact Riemann surfaces, since they become normalizable 
and are dynamical variables in the effective Lagrangian. 
More specifically, the standard moduli matrix on noncompact 
plane contains up to only $(k-1)$-th power of $z$ except 
in the first component. 
The $N_f-1$ coefficients of these $(k-1)$-th power represent 
``size'' of vortices, are nonnormalizable, and have to be 
fixed by the boundary condition on noncompact plane. 
Both of the ``vacuum'' and the ``size'' moduli become 
normalizable and provide additional $2N_f-2$ complex moduli 
on compact Riemann surfaces. 
 From now on, we take Eq.~(\ref{eq:u1moduliMatrix_kvortex}) 
as the general moduli matrix for the $k$-vortex sector. 

By inserting the moduli matrix (\ref{eq:u1moduliMatrix_kvortex}) 
 and the conformal factor (\ref{eq:confFactorSphere}) 
into Eq.~(\ref{eq:kahlerPotential}), we obtain the K\"ahler 
potential on the sphere 
\begin{eqnarray}
K^{(k)}={\cal A} c \int_{S^2} dx dy \frac{1}{\pi(1+|z|^2)} 
\log\left(\sum_{i=1}^{N_f}
\left|\sum_{j=0}^{k}a_j^{(i)}z^j\right|^2\right)
.
\label{eq:kahlerPotential_kortex}
\end{eqnarray}
We find that the integral is convergent and is proportional to 
${\cal A}$, indicating that all the moduli parameters take 
values of order ${\cal A}$. 
 Eqs.~(\ref{eq:effectiveLagrangianStrongCoupling}) 
and (\ref{eq:kahlerPotential_kortex}) give 
the effective Lagrangian of 
$k$ semi-local vortices on the sphere. 
 From the effective Lagrangian, we can define the 
metric in the moduli space 
\begin{eqnarray}
ds^2_{\rm mod}(S^2, k)=N {\cal A} c 
\sum_{i=1}^{N_f}\sum_{j=0}^{k}
\sum_{i'=1}^{N_f}\sum_{j'=0}^{k}
d{a}_j^{(i)} d{\bar a}_{j'}^{(i')} 
\frac{\partial}{\partial a_j^{(i)}} 
\frac{\partial}{\partial \bar a_{j'}^{(i')}} 
K^{(k)}.
\label{eq:metric_u1_kVortex}
\end{eqnarray}
So we find that each complex moduli gives a power of 
$\cal A$.
Since there are $kN_f+N_f-1$ complex moduli, we obtain 
the volume of the moduli space asymptotically ${\cal A}\to\infty$ 
to be proportional to the $kN_f+N_f-1$ power of ${\cal A}$ 
\begin{eqnarray}
\hat{\cal V}_{k}^{1, N_f}(S^2)&=&
\left(Nc\pi{\cal A}\right)^{kN_f+N_f-1}
\nonumber \\
&&
\times \int \prod_{i=1}^{N_f}\prod_{j=0}^{k}
d{a}_j^{(i)} 
\prod_{i'=1}^{N_f}\prod_{j'=0}^{k}
d{\bar a}_{j'}^{(i')} 
\frac{\partial}{\partial a_j^{(i)}} 
\frac{\partial}{\partial \bar a_{j'}^{(i')}} 
\left(\frac{K^{(k)}}{c\pi{\cal A}}\right), 
\label{eq:volMod_sphere_u1_kVortex}
\end{eqnarray}
where the coefficient of $(Nc\pi{\cal A})^{kN_f+N_f-1}$ 
is given by an integral representation for $K^{(k)}$. 
This asymptotic power agrees with the result 
 (\ref{eq:u1Nfvolume}) of the topological field 
theory.

\subsection{Moduli Space Metric of Non-Abelian 
Vortices}

Let us next work out the effective Lagrangian 
of non-Abelian semi-local vortices ($N_f>N_c>1$) 
on the sphere, using the strong 
coupling $g^2\to\infty$.

In the same spirit as the Abelian vortices, we should 
choose the moduli matrices to satisfy 
the boundary condition (\ref{eq:vorticity}). 
The maximal degree of all possible $N_c\times N_c$ minor 
determinants of $N_c\times N_f$ matrix $H_0$ should be $k$. 
Following the method of K\"ahler quotient\cite{Hitchin:1986ea}, 
we parametrize 
the moduli matrix as 
\begin{eqnarray}
H_0^{(k)}(z) =\sqrt{c}\left(\mathbb{D}(z), \mathbb{Q}(z)\right)
, 
\label{eq:unmoduliMatrix_kvortex} 
\end{eqnarray}
where $\mathbb{D}$ is an $N_c\times N_c$ matrix and 
$\mathbb{Q}$ is an $N_c\times \tilde N_c$ matrix 
with $\tilde N_c=N_f-N_c$. 
Let us define 
\begin{eqnarray}
P(z)=\det\mathbb{D}(z). 
\label{eq:determinantmoduliMatrix} 
\end{eqnarray}
Contrary to the usual parametrization\cite{Eto:2007yv}, 
we require that the degree of determinants of all possible 
$N_c\times \tilde N_c$ minor matrices of $H_0(z)$ to 
be at most $k$, instead of the usual $k-1$. 
If the base space is a flat non-compact plane, 
these extra moduli represent orientation of ``vacuum'' 
at infinity, and are non-normalizable and discarded. 
On compact Riemannn surfaces, however, we should retain 
these ``vacuum'' moduli, since they become normalizable. 
Moreover, we find that the above moduli matrix also 
contains the so-called ``size'' moduli in the usual 
parametrization of moduli matrix, which are nonnormalizable 
on noncompact plane. 
We find that not only the ``vacuum'' moduli, 
but also these ``size'' moduli become normalizable, and 
should be retained on compact Riemann surfaces, 
similarly to the case of Abelian gauge theories.

Following Ref.\citen{Eto:2007yv}, we define the 
following $N_c\times \tilde N_c$ matrix $\mathbb{F}$
\begin{eqnarray}
\mathbb{F}(z)=P(z)\mathbb{D}^{-1}\mathbb{Q}(z). 
\label{eq:FMatrix} 
\end{eqnarray}
We also define the following $N_c \times k$ matrices 
${\bf{\Phi}}$ and $\mathbb{J}$ 
\begin{eqnarray}
\mathbb{D}(z){\bf{\Phi}}(z)=\mathbb{J}P(z)=0, \quad 
{\rm mod} \quad P(z). 
\label{eq:PhiJMatrix} 
\end{eqnarray}
Multiplication of ${\bf{\Phi(z)}}$ by $z$ can be divided by 
a polynomial $P(z)$ and gives constant $k\times k$ matrix 
$\mathbb{Z}$ and $N_c\times k$ matrix ${\bf{\Psi}}$ 
\begin{eqnarray}
z{\bf{\Phi}}(z)={\bf{\Phi}}(z)\mathbb{Z}+{\bf{\Psi}}P(z), 
\label{eq:ZPsiMatrix} 
\end{eqnarray}
The matrices $\mathbb{Z}$ and ${\bf{\Psi}}$ gives moduli 
parameters. 
 From the definition, we obtain the relation 
\begin{eqnarray}
\mathbb{D}(z)\mathbb{F}(z)=P(z)\mathbb{Q}(z). 
\label{eq:FMatrix2} 
\end{eqnarray}
The matrix $\mathbb{F}$ now has a degree $k$ 
instead of the usual $k-1$, it is a linear combination of 
column vectors of ${\bf \Phi}$ after a division by $P(z)$ 
\begin{eqnarray}
\mathbb{F}(z)=\mathbb{A}P(z)+{\bf \Phi}(z)\tilde{\bf \Psi}, 
\label{eq:ABMatrix2} 
\end{eqnarray}
where constant $N_c\times\tilde N_c$ and $k\times \tilde N_c$ 
matrices $\mathbb{A}$ and $\tilde{\bf \Psi}$ are the remaining 
moduli parameters. 
In particular, the matrix $\mathbb{A}$ is a new moduli 
parameters which are discarded as non-normalizable 
if the base space is non-compact. 
Summarizing the moduli parameters in the moduli matrix 
(\ref{eq:unmoduliMatrix_kvortex}), we find 
$k\times k$ matrix $\mathbb{Z}$, 
$N_c\times k$ matrix ${\bf \Psi}$, $k\times \tilde{N}_c$ 
matrix $\tilde{\bf \Psi}$, and $N_c\times \tilde N_c$ 
matrix $\mathbb{A}$ as complex moduli parameters. 
Since the points in the moduli space are one-to-one
correspondence with the $GL(k, \mathbb{C})$ 
equivalence classes \cite{Eto:2007yv}, 
we find that the complex dimension of the moduli space 
of $k$-vortex sector is given by 
$kN_f+N_c(N_f-N_c)$.

By inserting the moduli matrix (\ref{eq:unmoduliMatrix_kvortex}) 
 and the conformal factor (\ref{eq:confFactorSphere}) 
into Eq.~(\ref{eq:kahlerPotential}), we obtain the K\"ahler 
potential on the sphere 
\begin{eqnarray}
K^{(k)}={\cal A} c \int_{S^2} dx dy \frac{1}{\pi(1+|z|^2)} 
\log\det
\left(|P(z)|^2+ \mathbb{F}(z)\mathbb{F}^\dagger(z)\right)
.
\label{eq:unkahlerPotential_kortex}
\end{eqnarray}
We find that the integral is convergent and is proportional to 
${\cal A}$, indicating that all the moduli parameters take 
values of order ${\cal A}$. 
 Eqs.~(\ref{eq:effectiveLagrangianStrongCoupling}) 
and (\ref{eq:unkahlerPotential_kortex}) give 
the effective Lagrangian of 
$k$ non-Abelian semi-local vortices on the sphere. 
Since there are $kN_f+N_c(N_f-N_c)$ complex moduli, we obtain 
the volume of the moduli space asymptotically ${\cal A}\to\infty$ 
to be proportional to the $kN_f+N_c(N_f-N_c)$ power of ${\cal A}$,  
in agreement with the result (\ref{eq:volumeGeneralNcNf}) 
of the topological field theory. 

If we let $N_f=N_c=N$, we obtain the so-called non-Abelian 
local vortices. 
In this case, the moduli matrix (\ref{eq:unmoduliMatrix_kvortex}) 
does not have a $\mathbb{Q}$ piece. 
For single vortex $k=1$, the metric has been 
obtained explicitly\cite{Fujimori:2010fk} 
and only the position moduli can 
be of order $\sqrt{\cal A}$, whereas other orientational 
moduli consists of $\mathbb{C}P^{N-1}$ with the radius 
of order $1/g\sqrt{c}$. 
 Moduli space of multi-vortices $k >1$ is symmetric 
product of $k$ moduli spaces of each single vortex 
except for separations of order smaller than the vortex 
size $1/g\sqrt{c}$. 
These facts imply that the orientational moduli can 
only give a finite volume unrelated to ${\cal A}$, 
whereas the vortex position can be of order $\sqrt{\cal A}$. 
Therefore the volume of the moduli space for $k$ local 
non-Abelian vortices is proportional to ${\cal A}^{k}$, 
which agrees with our result (\ref{eq:nLocalVortexVolume}) 
of the topological field theory.

\subsection{Moduli Space Metric at Finite Couplings}

In this subsection, we consider the effective Lagrangian 
at finite gauge couplings. 
For simplicity, we take Abelian semi-local vortices. 
Let us first take the vacuum sector $k=0$ with the 
moduli matrix 
\begin{eqnarray}
H_0^{(0)}(z) =\sqrt{c}(a^{(1)}, \cdots, a^{(N_f)}). 
\label{eq:moduliMatrix_0vortex} 
\end{eqnarray}
The corresponding solution of the master equation 
(\ref{eq:masterEqUn}) is given by $\Omega=H_0H_0^\dagger/c$, 
and we find $\partial_z(\Omega^{-1} \delta_t^\dagger \Omega)=0$. 
 By inserting this moduli matrix to 
Eq.~(\ref{eq:effectiveLagrangian1}), 
we obtain 
\begin{eqnarray}
L_{\rm eff}^{(k=0)}(S^2)
&=&
c \; \int_{S^2} dx dy \, \sigma \,
\delta_t^\dagger \left[
\frac{\delta_t\left(\sum_{j=1}^{N_f}|a^{(j)}|^2\right)}
{\sum_{l=1}^{N_f}|a^{(l)}|^2}
\right] 
\nonumber \\
&=& c \; {\cal A} \; 
\delta_t^\dagger \delta_t 
\log\left(\sum_{j=1}^{N_f}|a^{(j)}|^2\right). 
\label{eq:EffectiveLagrangianVacuum}
\end{eqnarray}
This is precisely the nonlinear sigma model with 
the target space $\mathbb{C}P^{N_f-1}$ of area $c{\cal A}$. 
The moduli $a^{(j)}$ are homogeneous coordinates of 
the K\"ahler manifold $\mathbb{C}P^{N_f-1}$ with 
the K\"ahler potential 
$\log\left(\sum_{j=1}^{N_f}|a^{(j)}|^2\right)$. 

Interpreting the effective Lagrangian in 
Eq.(\ref{eq:EffectiveLagrangianVacuum}), 
we can define the line element $ds_{\rm mod}^{(k=0)}(S^2)$ 
of the moduli space for the vacuum sector on the sphere as 
\begin{eqnarray}
ds^2_{\rm mod}(S^2, k=0)
&=&
N\; c\; {\cal A}
\left(\frac{da^{(i)} d\bar b^{(i)}}{\sum_{l=1}^{N_f} |b^{(l)}|^2}
-\frac{db^{(i)} \bar b^{(i)} b^{(j)} d\bar b^{(j)}} 
{(\sum_{l=1}^{N_f} |b^{(l)}|^2)^2}\right) , 
\label{eq:line_sphere_vacuum}
\end{eqnarray}
where we have put an arbitrary normalization factor $N$ 
in defining the moduli space metric from the effective Lagrangian.

The volume of the moduli space of the vacuum sector $k=0$ 
in the case of the sphere is given by 
\begin{eqnarray}
\hat{\cal V}_{k=0}^{1, N_f}(S^2)&=&
\left(Nc\pi{\cal A}\right)^{N_f-1}\frac{1}{(N_f-1)!} 
\label{eq:volMod_sphere_vacuum}
\end{eqnarray}

Let us next evaluate the moduli space metric 
of single vortex ($k=1$) on a sphere. 
It is not possible to obtain the metric analytically 
for generic point of the moduli space, even for the 
usual flat non-compact base space\cite{Eto:2007yv}. 
To understand the behavior of the moduli of a single 
vortex on a sphere near the Bradlow limit, 
we will first restrict ourselves to 
the moduli matrix where zeros of all components 
coincide 
\begin{eqnarray}
H_0^{(k=1)}(z) =\sqrt{c}(z-z_0) (a^{(1)}_1, \cdots, a^{N_f}_1), 
\label{eq:moduliMatrixU1Sphere} 
\end{eqnarray}
where the complex moduli parameters 
$(a^{(1)}_1, \cdots, a^{N_f}_1)$ 
are again inhomogeneous coordinates of $\mathbb{C}P^{N_f-1}$. 

For Abelian gauge theories, the master equation can be written 
in terms of $\psi\equiv \log \Omega$. 
The master equation (\ref{eq:masterEqUn}) on the 
sphere reads 
\begin{eqnarray}
\partial_z \partial_{\bar z} \psi=\frac{g^2 c}{4}
\left(1-e^{-{\psi}}(\sum_{j=1}^{N_f}|a^{j}_1|^2) 
|z-z_0|^2 \right)\sigma . 
\label{eq:masterEqU1single}
\end{eqnarray}
By defining a new function 
 $\psi^{(1)}$ 
\begin{eqnarray}
 \psi^{(1)}=\psi -\log(1+|\vec{b}|^2), 
\label{eq:ANOpsi}
\end{eqnarray}
we can rewrite the master 
equation (\ref{eq:masterEqU1single}) 
\begin{eqnarray}
\partial_z \partial_{\bar z} \psi^{(1)}=\frac{g^2 c}{4}
\left(1-e^{-\psi^{(1)}} |z-z_0|^2 \right)\sigma. 
\label{eq:masterEqU1singleANO}
\end{eqnarray}
since $(1+|\vec{b}|^2)$ is independent of the world-volume 
coordinates $z, \bar z$. 
This is precisely the same master equation for 
the single ANO vortex ($N_c=N_f=1$) on the Riemann surface. 
Moreover, $\psi$ in the boundary condition 
(\ref{eq:vorticity}) 
can be replaced 
by $\psi^{(1)}$, 
since $(1+|\vec{b}|^2)$ is independent of $z, \bar z$ 
\begin{eqnarray}
2\pi k 
= -i \oint dz  \partial_z \psi 
= -i \oint dz  \partial_z \psi^{(1)}. 
\label{eq:unit_vorticity}
\end{eqnarray}
From the geometric reason, $\psi^{(1)}$ 
should have a logarithmic singularity 
which depends on the geodesic distance between 
$z$ and $z_0$. 
This implies that the variational derivative with respect 
to the antiholomorphic moduli $\bar z_0$ can depend on $\zb$, 
but not on $z$. Terefore we obtain 
\begin{eqnarray}
 \partial_z(\Omega^{-1}\delta_t^\dagger \Omega)
= \partial_z \delta_t^\dagger \psi
= \partial_z \delta_t^\dagger (\psi^{(1)} +\log(1+|\vec{b}|^2))
= \partial_z \dot{{\bar z}}_0\frac{\partial \psi^{(1)}}{\partial \bar z_0} 
=0. 
\label{eq:vanishing_2ndterm2}
\end{eqnarray}
The effective Lagrangian 
(\ref{eq:effectiveLagrangian1}) with the moduli 
matrix (\ref{eq:moduliMatrixU1Sphere}) becomes 
\begin{eqnarray}
&&L_{\rm eff}+2\pi c =
c\; \int_{S^2} dx dy \, \sigma 
\delta_t^\dagger \left[e^{-\psi}
\delta_t\left\{(\sum_{j=1}^{N_f}|a^{j}_1|^2) |z-z_0|^2 \right\}
\right] 
\\
&=&
c\; \int_{D_0} dx dy \delta_t^\dagger 
\left[\sigma e^{-\psi}(\sum_{j=1}^{N_f}|a^{j}_1|^2)|z-z_0|^2
\left\{\frac{\delta_t(\sum_{j=1}^{N_f}|a^{j}_1|^2)}
{\sum_{j=1}^{N_f}|a^{j}_1|^2}
+
\frac{\delta_t (z-z_0)}
{\bar z-\bar z_0}
\right\}\right]
,\nonumber
\label{eq:1vortexEffectiveLagrangian1}
\end{eqnarray}
where we replaced the integration region by a region $D_0$ 
with an infinitesimal hole around the vortex position $z_0$, 
since we can safely ignore the contribution from the 
hole because of the smooth integrand without a singularity. 
By using the master equation (\ref{eq:masterEqU1single}), 
we obtain 
\begin{eqnarray}
L_{\rm eff}+2\pi c 
&=&
c\; \int_{D_0} dx dy \, \delta_t^\dagger 
\left[\left(
\sigma -\frac{4}{g^2c}\partial_z\partial_{\bar z}\psi
\right)
\left\{\frac{\delta_t(\sum_{j=1}^{N_f}|a^{j}_1|^2)}
{\sum_{j=1}^{N_f}|a^{j}_1|^2}
+
\frac{\delta_t (z-z_0)}
{\bar z-\bar z_0}
\right\}\right]
\nonumber \\
&=& L_1+L_2. 
\label{eq:EffectiveLagrangianL1L2}
\end{eqnarray}

Since there is no singularity in the integrand of $L_1$, 
we can take the limit of the vanishing size of the 
hole. 
By using the definition of area 
(\ref{eq:areaConformalfactor}) and 
vortex number (\ref{eq:vorticity}) with $k=1$, 
we readily find 
\begin{eqnarray}
L_1
&=&
\delta_t^\dagger 
\left[\frac{\delta_t(\sum_{j=1}^{N_f}|a^{j}_1|^2)}
{\sum_{j=1}^{N_f}|a^{j}_1|^2}
\int_{S^2}
dx dy \left(
c\sigma -\frac{4}{g^2}\partial_z\partial_{\bar z}\psi
\right)
\right]
\nonumber \\
&=&
\left(
c{\cal A} -\frac{4\pi}{g^2}
\right)
\left(\frac{\dot{a}^{(i)}_1 \dot{\bar a}^{(i)}_1}
{\sum_{j=1}^{N_f}|a^{j}_1|^2}
-\frac{\dot{a}^{(i)}_1 \bar a^{(i)}_1 a^{(j)}_1 \dot{\bar a}^{(j)}_1} 
{\sum_{j=1}^{N_f}|a^{j}_1|^2}\right) . 
\label{eq:L1result}
\end{eqnarray}
This is precisely the standard metric of 
$\C P^{N_f-1}$ orientational moduli $\vec{b}$ 
with the radius $\sqrt{c{\cal A} -\frac{4\pi}{g^2}}$. 


On the other hand, the integrand of $L_2$ can be put into 
a total derivative 
\begin{eqnarray}
L_2
&=&
-\frac{4}{g^2}
\int_{D_0} dx dy \partial_{\bar z}
\left[
\frac{\delta_t (z-z_0)}
{z-z_0}
\delta_t^\dagger 
\partial_z\psi
\right]
\nonumber \\
&=&\frac{4}{g^2}
\int_{\partial D_0} \frac{dz}{2i} 
\frac{\delta_t (z-z_0)}
{z-z_0}
\delta_t^\dagger 
\partial_z\psi
\nonumber \\
&=&-\frac{4\pi}{g^2}\dot{ z}_0 
\left[\delta_t^\dagger \partial_z 
\psi\right]_{z=z_0},
\label{eq:L1Residue}
\end{eqnarray}
where the contour integration is around $z=z_0$ 
and is counter-clock-wise, since the sphere 
has no boundary and there is only a hole around 
$z=z_0$. 
We can replace $\psi$ in the effective Lagrangian 
(\ref{eq:L1Residue}) by $\psi^{(1)}$, 
since $(1+|\vec{b}|^2)$ is independent of $z, \bar z$ 
\begin{eqnarray}
L_2
&=&-\frac{4\pi}{g^2}\dot{ z}_0 
\left[\delta_t^\dagger \partial_z 
\psi^{(1)}
\right]_{z=z_0} 
.
\label{eq:L1Residue2}
\end{eqnarray}
Therefore we find that $L_2$ is identical 
to the effective Lagrangian for 
the single ANO vortex ($N_c=N_f=1$) on the Riemann surface.

We can evaluate the residue 
$[\delta_t^\dagger \partial_z \psi^{(1)}]_{z=z_0}$ 
by expanding around $z=z_0$ the solution of the 
master equation $\psi^{(1)}$ for the ANO vortex 
\begin{eqnarray}
\psi^{(1)}(z, \bar z) 
&=&-\log|z-z_0|^2
-a 
-\frac{\bar b}{2}(z-z_0) 
-\frac{ b}{2}(\bar z-\bar{z}_0) 
\nonumber \\
&&+\frac{g^2 c}{4}\sigma(z_0, \bar{z}_0) |z-z_0|^2 
+\cdots 
, 
\label{eq:expansionPsi}
\end{eqnarray}
where the coefficient of $|z-z_0|^2$ is determined by 
the master equation. 
We need to obtain the moduli dependence of the 
coefficient \cite{Samols:1991ne} 
\begin{eqnarray}
\left[\delta_t^\dagger \partial_z 
\psi^{(1)}
\right]_{z=z_0} 
=\dot{\bar{z}}_0
\left(-\frac{g^2 c}{4}\sigma(z_0, \bar{z}_0)
-\frac{1}{2}\frac{\partial \bar b}{\partial \bar{z}_0} 
\right) .
\label{eq:samolsFormula}
\end{eqnarray}
Since the geodesic distance is the same as the chordal 
distance $2|z-z_0|/\sqrt{(1+|z|^2)(1+|z_0|^2)}$ near $z \approx z_0$ 
up to the order that we are interested in, 
we obtain the moduli dependence of the expansion 
coefficients $b, \bar b$ as 
\begin{eqnarray}
\psi^{(1)}(z, \bar z) 
&=&-\log \left(\frac{|z-z_0|^2(1+|z_0|^2)}{1+|z|^2}\right) 
\nonumber \\
&=&-\log|z-z_0|^2
+ \frac{{\bar z}_0}{1+|z_0|^2}(z-z_0)
+ \frac{ z_0}{1+|z_0|^2}(\bar z-\bar{z}_0)
+\cdots 
, 
\label{eq:expansionHatPsi}
\end{eqnarray}
By comparing the expansion with Eqs.(\ref{eq:expansionPsi}) 
and (\ref{eq:expansionHatPsi}), we find 
\begin{eqnarray}
{\bar b}
=- 2\frac{{\bar z}_0}{1+|z_0|^2}, 
\quad
b
=
-2 \frac{ z_0}{1+|z_0|^2}
.
\label{eq:besphere}
\end{eqnarray}
Thus we find $L_2$ on sphere as 
\begin{eqnarray}
L_2(S^2)
=c
\left({\cal A}- \frac{4\pi}{g^2c}\right) 
\frac{|\dot{z}_0|^2}{(1+|z_0|^2)^2}
\label{eq:L2resultSphere}
\end{eqnarray}
Combining this with Eq.(\ref{eq:L1result}), 
we can now define the line element $ds_{\rm mod}$ 
in the case of vortex on the sphere as 
\begin{multline}
ds^2_{\rm mod}(S^2, k=1)
=\\
Nc\left({\cal A}- \frac{4\pi}{g^2c}\right)
\left[ \frac{|dz_0|^2}{(1+|z_0|^2)^2} 
+
 \left(\frac{db^i d\bar b^i}{1+ |\vec{b}|^2}
-\frac{db^i \bar b^i b^j d\bar b^j} 
{(1+|\vec{b}|^2)^2}\right)
\right]. 
\label{eq:line_sphere}
\end{multline}
Although the moduli matrix covers only a part of moduli 
space, the result shows that both the vortex position $z_0$ 
and orientational moduli $\vec{a}_1$ are meaningful 
only for ${\cal A} \ge 4\pi/(g^2c)$, satisfying the Bradlow 
bound. 
It is interesting to note that the same $\mathbb{C}P^{N_f-1}$ 
orientational moduli as the vacuum moduli emerges except 
that their radius is changed from $\sqrt{\cal A}$ 
to  $\sqrt{{\cal A} - 4\pi/(g^2c)}$ satisfying 
the the Bradlow bound. 



\section{Comparison with Hanany-Tong Moduli Space}

So far, we have discussed the volume of the moduli space 
of the BPS equations for the vortex. 
According to Ref.~\citen{Hanany:2003hp}, the string 
theoretical realization of the BPS vortex admits 
an alternative description of the vortex moduli space. 
Hanany and Tong have defined the 
ADHM-like 
quotient space 
from effective theory on the brane configuration 
for vortex in superstring theory and discussed the 
properties of the moduli space of the vortex. 
The volume of the moduli space corresponding 
to the Hanany-Tong (HT) quotient space has already 
evaluated by the localization method 
\cite{Moore:1998et,Shadchin:2006yz,Ohta:2007ji,Dimofte:2010tz,Awata:2010bz,Yoshida:2011au,Bonelli:2011fq}.
In this section, we extend these evaluations to the 
non-Abelian case and compare with the results in the 
previous sections.

Using the brane configuration of the supersymmetric gauge 
theory, the $2+1$-dimensional gauge theory is realized on 
the D3-branes stretched between two NS5-branes.
The world-volumes of the D3-branes and NS5-branes are along 
$(x^0,x^1,x^2,x^6)$ and $(x^0,x^1,x^2,x^3,x^4,x^5)$ 
directions, respectively. The rank of the gauge theory 
corresponds to the number of the D3-branes and the inverse 
of the square of the gauge coupling $1/g^2$ is proportional 
to the distance of two NS5-branes. 
We can also add the matter fields (hypermultiplets) in the 
fundamental representation by introducing semi-infinite 
D3-branes from one side of the NS5-branes. 
The number of the semi-infinite D3-branes is the number of 
the flavors $N_f$.

In the Higgs phase, the D3-branes between the NS5-branes 
connects with the semi-infinite D3-branes. 
Then two NS5-branes can move relatively along 
$(x^7,x^8,x^9)$-directions. The distance of the NS5-branes 
along these directions relates to the FI-parameters of the 
gauge theory. 
Turning on the FI-parameters, the vortex is admitted in 
this system. 
The $k$ vortices are described by 
$k$ D1-branes stretching between the NS5-brane and the 
connected D3-branes. (See Fig.~1.)

\begin{figure}[t]
\begin{center}
\begin{tabular}{ccc}
\includegraphics[scale=0.38]{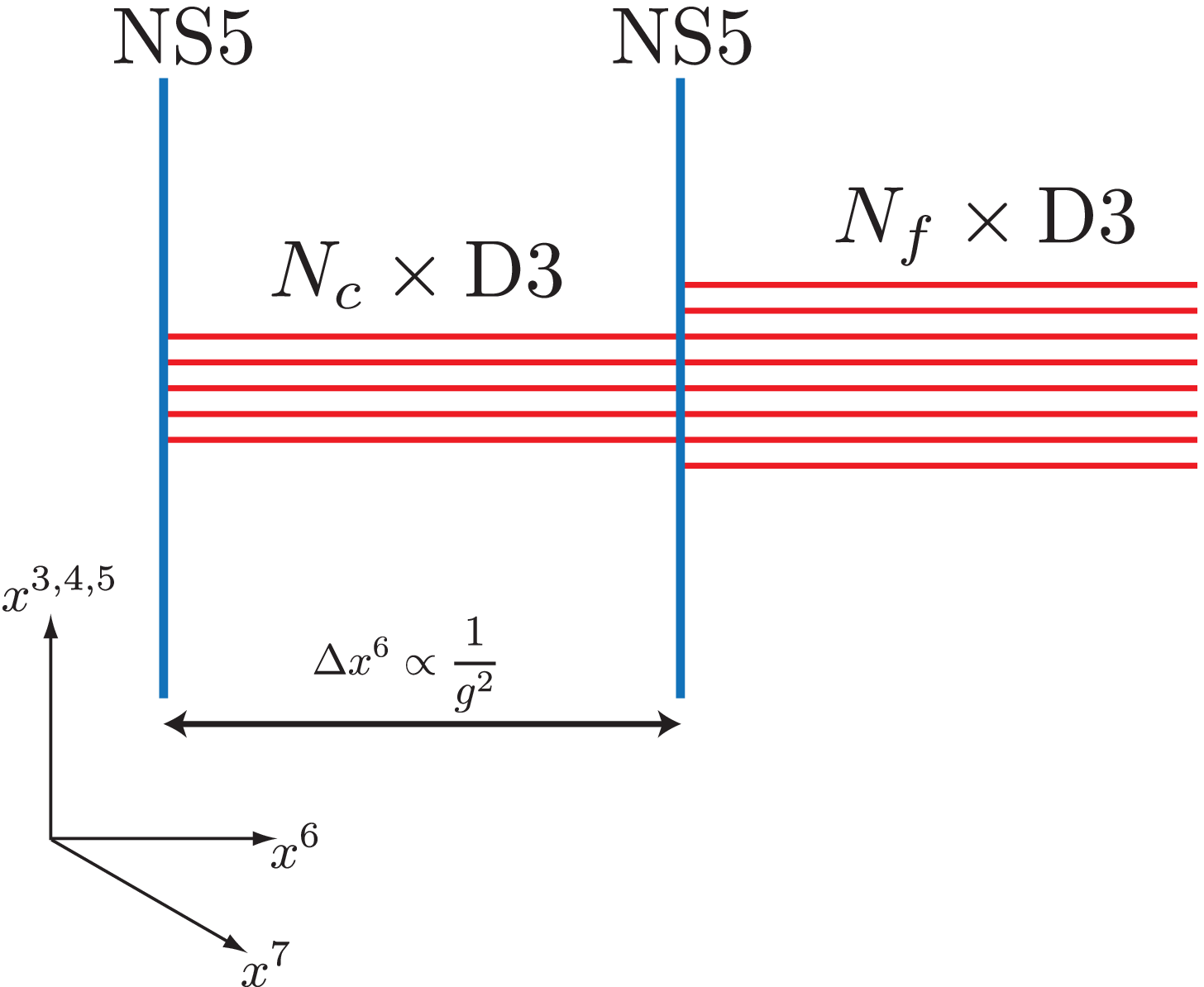}
&&
\includegraphics[scale=0.38]{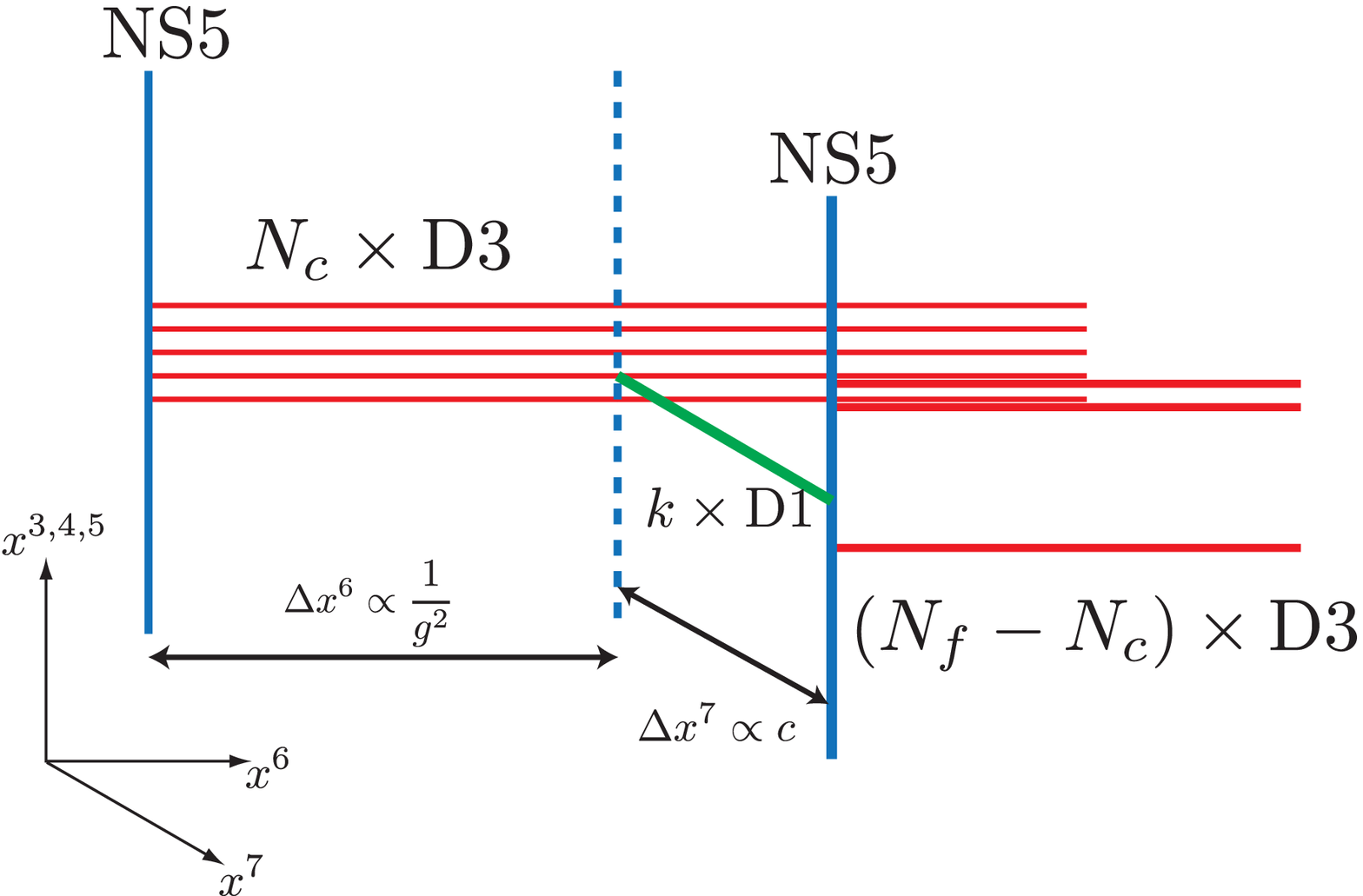}\\
(a) && (b)
\end{tabular}
\end{center}
\caption{Brane configuration of 3-dimensional 
supersymmetric gauge theory in Higgs phase.
(a) In the Higgs phase without the FI-parameter, 
$N_c$ D3-branes connect with semi-infinite (flavor) D3 branes.
(b) Turning on the FI-parameter $c$, the right NS5-brane 
moves along $x^7$-direction 
and $k$ D1-branes between $N_c$ D3-branes and NS5-brane 
with $\tilde{N}_c=N_f-N_c$ semi-infinite D3-branes 
appear as the vortices in the Higgs phase.}
\end{figure}

Hanany and Tong have considered that the moduli space 
of the vortex is realized by the vacuum equations 
of the effective theory on the D1-branes. 
The effective theory on the $k$ D1-branes is  
a $U(k)$ gauge theory with $N_c$ 
chiral superfields in the fundamental $k$ representation 
and $\tilde{N}_c\equiv N_f-N_c$ chiral superfields in the 
anti-fundamental $\bar k$ representation. 
From the brane configuration, we can see 
that the role is exchanged between the gauge coupling 
and the FI-parameter, when we consider the effective 
$U(k)$ gauge theory on the vortex instead of the 
$U(N_c)$ gauge theory, and vice versa. 
So the FI-parameter $r$ of the effective $U(k)$ theory is 
proportional to $1/g^2$. 
Strictly speaking, the effective theory on the D1-branes 
is 
a $1+1$-dimensional gauge theory. 
However 
we reduce the effective theory to a $0$-dimensional 
theory (matrix model), 
 since we are here interested in (vacuum) solutions of 
the effective theory that are independent of 
the world-volume coordinates on the D1-branes.

Considering the vacuum equation of the reduced effective 
theory on the D1-branes, 
the HT moduli space is defined by the \Kahler 
quotient of the following moment map
\be
\mu_{\text{HT}} = [Z,Z^\dag] + II^\dag-J^\dag J -r,
\label{HT moment map}
\ee
where $Z$, $I$ and $J$ are $k\times k$, $k\times N_c$, 
and $\tilde{N}_c \times k$  matrices, respectively. 
The quotient space is
\be
\M_k^{\text{HT}} =  \frac{\mu_{\text{HT}}^{-1}(0)}{U(k)}.
\ee

Introducing fermionic partners of $Z, I, J$ as 
$\lambda, \psi_I, \psi_J$, we define the BRST 
transformations for these matrices  
\be
\begin{array}{lcl}
Q Z = i\lambda, && Q \lambda = -[\Phi,Z],\\
Q Z^\dag = -i\lambda^\dag, && Q \lambda^\dag = [\Phi,Z^\dag],\\
Q I = i\psi_I, && Q\psi_I = \Phi I-IM,\\
Q I^\dag = -i\psi_I^\dag, && Q \psi_I^\dag = I^\dag \Phi -MI^\dag,\\
Q J = i\psi_J, && Q\psi_J = J \Phi - \tilde{M}J,\\
Q J^\dag = -i\psi_J^\dag, && 
Q \psi_J^\dag =  \Phi J^\dag - J^\dag\tilde{M},\\
Q \Phi = 0,&& \\
\end{array}
\ee
where $\Phi$ is now $k\times k$ matrix. 
Note that $\Phi$ differs from that in the previous sections.
We assume for a while the masses for the hypermultiplets 
$I$ and $J$, $M=\diag.(m_1,m_2,\ldots,m_{N_c})$ and 
$\tilde{M}=\diag.(m_1,m_2,\ldots,m_{\tilde{N}_c})$, 
are not 
degenerate for simplicity of the calculations 
while so far we have treated the degenerate 
cases. 

Following the same argument in the previous sections, 
the volume of the HT moduli space $V_k^{N_c,N_f}$ is 
given by the following matrix integral 
\be
V_k^{N_c,N_f}(\vec{m},\vec{\tilde{m}},r) =  
\int \D \Phi  \D^2 Z \D^2 \lambda \D^2 I \D^2 J  
\D^2 \psi_I \D^2 \psi_J 
e^{-S},
\ee
where $S$ is a matrix model ``action'', which gives the 
constraint $\mu_{\text{HT}}=0$, defined by 
\be
S = \Tr \left[ i\Phi ([Z,Z^\dag]+II^\dag - J^\dag J-r)
-iI M I^\dag+iJ^\dag \tilde{M}J
+ \lambda \lambda^\dag -\psi_I \psi_I^\dag - \psi_J^\dag \psi_J
\right].
\ee
We can see the action is BRST closed, $QS=0$.

The solution $\mu_{\text{HT}}$ of the moment map 
(\ref{HT moment map}) contains the flat directions 
along the commuting elements of $Z$.
So the HT moduli space is non-compact and  its volume 
$V_k^{N_c,N_f}$ diverges in general.
To regularize the volume of the HT moduli space, we 
introduce the so-called 
$\Omega$-background.\cite{Moore:1997dj,Nekrasov:2002qd}
The $\Omega$-background deforms the BRST transformations as
\be
\begin{array}{lcl}
Q_\e Z = i\lambda, && Q_\e \lambda = -[\Phi,Z]-\e Z,\\
Q_\e Z^\dag = -i\lambda^\dag, && 
Q_\e \lambda^\dag = [\Phi,Z^\dag]-\e Z^\dag,\\
Q_\e I = i\psi_I, && Q_\e\psi_I = \Phi I-IM,\\
Q_\e I^\dag = -i\psi_I^\dag, && 
Q_\e\psi_I^\dag = I^\dag \Phi - MI^\dag,\\
Q_\e J = i\psi_J, && Q_\e\psi_J = J \Phi- \tilde{M}J+\e J,\\
Q_\e J^\dag = -i\psi_J^\dag, && 
Q_\e\psi_J^\dag =  \Phi J^\dag - J^\dag\tilde{M}+\e J^\dag,\\
Q_\e \Phi = 0,&& \\
\end{array}
\ee
where $\e$ is a parameter of the $\Omega$-background.
Note that the hypermultiplet masses $M$ and $\tilde{M}$ 
are independent of $\e$ 
in comparison with the instanton counting (ADHM equations),
where  there exist extra \HK constraints.

The action is also deformed to 
\begin{multline}
S_\e = \Tr \Big[ i\Phi ([Z,Z^\dag]+II^\dag - J^\dag J-r) +i\e ZZ^\dag
-iI M I^\dag+iJ^\dag (\tilde{M}-\e)J\\
+ \lambda \lambda^\dag -\psi_I \psi_I^\dag - \psi_J^\dag \psi_J
\Big].
\end{multline}
The parameter $\e$ appears as the mass term for $Z$ which  
lifts  
the flat directions of the constraint $\mu_{\text{HT}}=0$.
 
The 
regularized volume of the HT moduli space is  
\be
V^{N_c,N_f}_k(\vec{m},\vec{\tilde{m}},r;\e) =  \int \D \Phi  \D^2 Z \D^2 \lambda \D^2 I \D^2 J  \D^2 \psi_I \D^2 \psi_J  e^{-S_\e}.
\ee
After integrating all matrices except for $\Phi$, we obtain
\be
V^{N_c,N_f}_k(\vec{m},\vec{\tilde{m}},r;\e)  = \int \D \Phi \frac{1}{P(\Phi)Q(\Phi)}\frac{1}{\det(-i[\Phi,\cdot]+i\e)}e^{ir \Tr \Phi},
\ee
where $P(\Phi)=\det(i\Phi\otimes 1-i1\otimes M)$ and $Q(\Phi)=\det(i1\otimes \tilde{M}-i\Phi\otimes 1-i\e 1\otimes 1)$.
Then, diagonalizing $\Phi = \diag.(\phi_1,\phi_2,\ldots,\phi_k)$, we obtain
\begin{multline}
V^{N_c,N_f}_k(\vec{m},\vec{\tilde{m}},r;\e)   = \frac{1}{i^{kN_f}\e^k}\int \prod_{a=1}^k \frac{d \phi_a}{2\pi i}
 \frac{1}{\prod_{a=1}^k\prod_{i=1}^{N_c} (\phi_a-m_i)\prod_{a=1}^k\prod_{i=1}^{\tilde{N}_c} (\tilde{m}_i-\phi_a-\e)}\\
 \times
 \prod_{a<b}\frac{(\phi_a-\phi_b)^2}{(\phi_a-\phi_b)^2-\e^2}e^{ir\sum_{a=1}^k \phi_a}.
 \label{HT integral}
\end{multline}

This integral is essentially the residue integral and localized at
the fixed points of the BRST transformations.
The fixed points is classified by $N_f$  
partitions of $k$, 
and the eigenvalues
of $\Phi$ at the fixed points are given by
\be
\Phi_0=
\begin{pmatrix}
m_1{\bf 1}_{k_1} + {\cal J}^{(k_1)} & & & & &\\
 & \ddots & & & &\\
 & & m_{N_c}{\bf 1}_{k_{N_c}} + {\cal J}^{(k_{N_c})}  & & &\\
 & & & \tilde{m}_1{\bf 1}_{\tilde{k}_1} + {\cal J}^{(\tilde{k}_1)}  & &\\
 & & & & \ddots &\\ 
 & & & & & \tilde{m}_{\tilde{N}_c}{\bf 1}_{\tilde{k}_1} + {\cal J}^{(\tilde{k}_{\tilde{N}_c})}
\end{pmatrix},
\ee
where $\sum_{l=1}^{N_c} k_l + \sum_{l'=1}^{\tilde{N}_c} k_l' = k$ and
${\cal J}^{(k)}$ is a $k\times k$ diagonal matrix which has the form
\be
{\cal J}^{(k)} =
\begin{pmatrix}
0 & & & & \\
 & \e & & & \\
 & & 2\e & & \\
 & & & \ddots & \\
 & & & & (k-1)\e
\end{pmatrix}.
\ee

Using this fixed point data, we can evaluate the integral (\ref{HT integral}).
To evaluate the integral, it is useful to introduce the ``character'' \cite{Nakajima:1999,Bruzzo:2002xf}

\be
\chi^{N_c,N_f}_k(t)
 =  (t-1) V\times V^* + W\times V^* + t V \times  \tilde{W}^*,
\ee
where $t=e^{-\beta\e}$, $V=\Tr e^{-\beta\Phi_0}$, $V^*=\Tr e^{\beta\Phi_0}$, $W=\Tr e^{-\beta M}$
 and $\tilde{W}^*=\Tr e^{\beta\tilde{M}}$.
This character is defined from the 
integrand (measure) of (\ref{HT integral})
by a K-theoretic extension.
The term $(t-1) V\times V^*$ corresponds to the Vandermonde difference products in (\ref{HT integral})
and the terms $W\times V^*$ and $ V \times  \tilde{W}^*$ come from the poles after integrating the hypermultiplets.
The 
volume
is obtained in the following procedure.
Once we obtain the character in polynomial of $t$
\be
\chi^{N_c,N_f}_k(t) = \sum_{i=1}^N a_i(m,\tilde{m};\beta) t^{n_i},
\ee
the volume 
is obtained in the limit of the
extra parameter $\beta$
\be
V^{N_c,N_f}_k(\vec{m},\vec{\tilde{m}},r;\e) 
=\lim_{\beta\to 0}\beta^d
\prod_{i=1}^N 
\frac{1}{1-a_i(m,\tilde{m};\beta) t^{n_i}}e^{ir\Tr\Phi_0},
\ee
where $d$ is the 
complex 
dimension of the moduli space.

The character replaces the products in the integrand of the partition
function (\ref{HT integral}) with the polynomials of $t$.
This evaluation of the character regularizes the residue integral 
and we can evaluate the partition function efficiently.


\koumoku{ANO vortex}

Let us first consider the simplest ANO vortex case, namely 
$k$ vortices of $N_c=N_f=1$. In this case, there is only 
one mass parameter $m$ and a 
single partition of $k$. 
$\Tr \Phi_0$ is rather simple
\be
\Tr \Phi_0 = \sum_{l=1}^k (m+(l-1)\e) = km+\e\frac{k(k-1)}{2},
\ee
and
$V$ and $V^*$ are given by
\be
V=e^{-\beta m}\frac{1-t^{k}}{1-t}, \qquad V^* =e^{\beta m}\frac{1-t^{-k}}{1-t^{-1}}.
\ee
If we evaluate the character, the mass dependence disappears
\be
\begin{split}
\chi^{1,1}_{k}(t) &= (t-1)e^{-\beta m}\frac{1-t^{k}}{1-t} \times e^{\beta m}\frac{1-t^{-k}}{1-t^{-1}}
+e^{-\beta m}\times e^{\beta m}\frac{1-t^{-k}}{1-t^{-1}}\\
&=t\frac{1-t^{k}}{1-t} =t +t^2 +t^3+\cdots +t^k.
\end{split}
\ee
From this character polynomial in $t$, we find the partition function
(regularized volume)
\be
\begin{split}
V^{1,1}_k(m,r;\e)  &=\lim_{\beta \to 0}\frac{\beta^k}{i^k} \prod_{i=1}^k \frac{1}{1-t^i}e^{ir(m k+\e k(k-1)/2)}\\
&=\frac{1}{(i\e)^k k!}e^{ir (km+\e k(k-1)/2)}.
\end{split}
\ee
To compare with our previous result, we must set $m=0$.
Then we obtain
\be
V^{1,1}_k(r;\e) 
=\frac{1}{(i\e)^k k!}e^{ir\e k(k-1)/2}.
\ee
This volume diverges in the $\e\to 0$ limit as expected.
The divergence should come from the position moduli of the vortices
since the commuting parts of the matrix $Z$ of the 
HT moment map 
(\ref{HT moment map}) 
correspond to the positions. In the $\e\to 0$ limit, 
the volume behaves as
\be
V^{N_c,N_f}_k(r;\e) \sim \frac{1}{(i\e)^k k!}.
\ee
On the other hand, our previous result  
(\ref{eq:asymptoticVolume}) 
also behaves as
\be
\calV^{N_c,N_f}_k \sim \frac{(2\pi\A)^k}{k!}.
\ee
In this observation, we notice that the asymptotic behavior agrees 
with each other by identifying $\A \sim \frac{1}{i\e}$.
Of course, the relation between the area of the Riemann surface $A$ and
the $\Omega$-background parameter $\e$ is unclear,
but both parameters regularize the divergences coming from the non-normalizable
modes of the moduli. So we naively expect that the asymptotic (divergent) behaviors of the 
volume of the HT moduli space and the moduli space of the vortices on
the compact Riemann surface coincide with each other, although the whole structures of the both
moduli spaces may differ.
To understand precise relation between two moduli spaces, we need further investigation
of the details of the moduli space, but we will check some examples to compare with the divergent behavior
of the volume of the HT moduli space in the following.

\koumoku{Local vortex}

Next we consider the case of $N_c=N_f=N>1$, where the 
solution is called the local vortex. 
In this case, the matrix $J$ disappears since $\tilde{N}_c =0$. 
Then the HT moment map 
(\ref{HT moment map}) 
becomes simpler
\be
\mu_{\text{HT}} = [Z,Z^\dag] + II^\dag-r.
\ee
The poles in (\ref{HT integral}) coming from the 
contribution of $J$ also disappears and the localization 
fixed points are classified by an $N$ partition of $k$, 
namely $\vec{k} = (k_1,k_2,\ldots,k_N)$. 
Using the fixed point data, we find 
\bea
&& \Tr \Phi_0=\sum_{l=1}^N \left[k_lm_l
+\e\frac{k_l(k_l-1)}{2}\right]\\
&&V=\sum_{l=1}^N e^{-\beta m_l}\frac{1-t^{k_l}}{1-t},
\quad
V^*=\sum_{l'=1}^N e^{\beta m_{l'}}\frac{1-t^{-k_{l'}}}{1-t^{-1}}.
\eea
Using those, the character is given by
\be
\begin{split}
\chi^{N,N}_k(t)  &= \sum_{l,l'=1}^N e^{-\beta (m_l-m_{l'})}\left(
(t-1)\frac{1-t^{k_l}}{1-t}\frac{1-t^{-k_{l'}}}{1-t^{-1}}
+\frac{1-t^{-k_{l'}}}{1-t^{-1}}
\right)\\
&=\sum_{l,l'=1}^N e^{-\beta (m_l-m_{l'})}
\sum_{j=1}^{k_{l'}}
t^{k_l-j+1}.
\end{split}
\ee
The volume of the HT moduli space becomes
\begin{multline}
V^{N,N}_k(\vec{m},r;\e) = 
\sum_{\sum_{l=1}^N k_l =k}
\frac{1}{\prod_{l,{l'}=1}^N \prod_{j=1}^{k_{l'}} 
(m_l-m_{l'}+\e(k_l-j+1))}\\
 \times e^{ir \sum_{l=1}^N (k_l m_l+\e k_l(k_l-1)/2)}.
 \label{HT volume of local vortex}
\end{multline}

To compare with our previous results, we have to take 
the limit $m_l \to 0$, but it is difficult to evaluate 
this limit for the non-Abelian local vortex in general.  
However, for the case of $k=1$, 
we can perform the integral (\ref{HT integral}) directly. 
Using the residue integral formula (\ref{residue integral}), 
\be
V^{N,N}_1(\vec{0} ,r;\e)=
\frac{1}{i\e}\int_{-\infty}^{\infty} \frac{d \phi}{2\pi} 
 \frac{1}{(i\phi)^N}
e^{ir\phi}=\frac{1}{i\e}\frac{r^{N-1}}{(N-1)!}. 
\ee
For general $k$, it is difficult to 
calculate the integral for fixed values of $\epsilon r$ 
in the limit of degenerate masses 
$m_l \to 0$. 
However if we instead fix $m_l$ and take the leading 
term of (\ref{HT volume of local vortex}) 
at the $\e\to 0$ limit which corresponds 
to the large area limit, we find the volume $V_k^{N,N}$ 
 is proportional to $1/(i\e)^k$ 
\be
\lim_{\e\to 0}V^{N,N}_k(\vec{m},r;\e) = \frac{1}{i^{k N}(i\e)^k}
\sum_{\sum_{l=1}^N k_l =k}
\left(\prod_{l=1}^{N} \frac{1}{k_l!}\right)
\frac{1}{\prod_{l\neq {l'}} (m_l-m_{l'})^{k_{l'}}} 
e^{ir \sum_l k_l m_l}. 
 \label{HT volume of local vortex limit}
\ee
If we next take the degenerate mass limit $m_l \to 0$, 
the coefficient of $1/(i\e)^k$ in 
(\ref{HT volume of local vortex limit}) should be 
proportional to $r^N$ from dimensional reasons.
The divergent behavior (divergent power) looks like the asymptotic volume of
the non-Abelian local vortex on the sphere (\ref{eq:nLocalVortexVolume})
in the large area limit,
but 
the coefficients depending on $N$ and $k_l$
 does not agree with the asymptotic behavior  
neither on the sphere (\ref{eq:nLocalVortexVolume}) nor on the torus
(\ref{large A on torus}) in \S4.
Since our treatment of Hanany-Tong moduli space does not carry 
informations of topology of base space, it is not clear at 
present how to relate the results of Hanany-Tong moduli space 
and those in previous sections. 


\koumoku{Semi-local vortex}

The semi-local vortex case ($N_f>N_c$) is more complicated.
At least, we can see from the integral 
(\ref{HT integral}) for the $k=1$ case
\be
\hat{\calV}^{N_c,N_f}_1(\vec{0},\vec{0},r;\e)   = \frac{1}{i\e}\int_{-\infty}^{\infty}  \frac{d \phi}{2\pi}
 \frac{1}{(i\phi)^{N_c}(-i\phi-i\e)^{\tilde{N}_c}}
 e^{ir \phi},
\ee
where we have set $m_i=\tilde{m}_i=0$. 
Using the residue formula, we find  that  
the volume of the HT moduli space behaves in general
\begin{multline}
\hat{\calV}^{N_c,N_f}_1(\vec{0},\vec{0},r;\e) =
\frac{1}{(-1)^{\tilde{N}_c}(i\e)}\Bigg[\sum_{l=1}^{N_c}\frac{(-1)^{N_c-l}(N_f-l-1)!}{(\tilde{N}_c-1)!(l-1)!(N_c-l)!}
\frac{r^{l-1}}{(i\e)^{N_f-l}}\\
+\sum_{l=1}^{\tilde{N}_c}\frac{(-1)^{\tilde{N}_c-l}(N_f-l-1)!}{(N_c-1)!(l-1)!(\tilde{N}_c-l)!}
\frac{r^{l-1}}{(i\e)^{N_f-l}}
e^{-i\e r}\Bigg].
\end{multline}
In the limit of $\e\to 0$, the divergent power of $1/(i\e)$ is $N_f$, which
agrees with the dimension of the moduli space of the vortices
 including the non-normalizable modes.

For general $k$, we expect that the volume of the HT moduli space
behaves $\hat{\calV}^{N_c,N_f}_k(\vec{0},\vec{0},r;\e)\sim \frac{1}{(i\e)^{kN_f}}$
in the $\e\to 0$ limit, but it is difficult to prove it.
We leave in the future deeper investigation of the volume of the HT moduli space
to understand the relation to our results.

\section{Conclusion and Discussion}

In this paper, we have investigated the volume of the 
moduli space of the Abelian/non-Abelian vortex on the 
compact Riemann surface by using the path integral in 
topological field theory. 
We can obtain 
the properties of the volume of the 
moduli space without any knowledge on the metric of the 
moduli space. 
This advantage of 
our 
field theoretical method 
makes easy to evaluate the volume for any number of the 
colors or flavors, and  
for 
any genus of the Riemann surface. 
The volume obtained by our method gives us much information on 
the moduli space, 
for example, dimension of the normalizable or 
non-normalizable modes, criterion of existence of 
the vortices, behavior near the Bradlow limit, 
and so on.

The metric on the moduli space can be defined from 
the effective Lagrangian on vortices. 
We have obtained the effective Lagrangian on the Riemann 
surface of sphere topology in the strong 
coupling limit $g^2\to \infty$, or for a restricted class of 
vortex number, $k=0, 1$ at finite couplings. 
By integrating the metric, we obtain the volume which 
agrees with our results of the topological field theory. 
Although more detailed information than the volume alone 
can be obtained from the effective Lagrangian, 
it is generally difficult to evaluate it explicitly. 
For instance, we have not understood the difference of the 
effective Lagrangian due to the topology 
of the Riemann surface, such as torus, or higher genus. 
We stress that our method of topological field theory 
is much more versatile than the method of the effective 
Lagrangian, and still can give valuable information such 
as the volume of the moduli space. 
Even without the knowledge of the metric, 
the volume from the field theoretical method may 
lead to a novel knowledge on the moduli space, 
which can 
not be captured from the effective Lagrangian.

The topological filed theory, which we have utilized for 
the evaluation of the volume, might be embedded in  
a supersymmetric gauge theory. Indeed,  
the BRST algebra in 
\S3 and \S4 could be constructed from the supercharges 
in two-dimensional ${\cal N}=(2,2)$ 
supersymmetric gauge theory by topological twisting. 
The cohomological 
operators correspond to
the supersymmetric (SUSY invariant) operators, and the 
``action'' is the essential part of the action of the 
supersymmetric gauge  
theory, 
in the sense of the localization.
Thus the volume evaluated in the path integral can be 
understood as the  
vacuum expectation value (vev) 
of the supersymmetric operator. 
This strongly suggests that our exact evaluation of the 
volume gives  an 
exact evaluation of the vev including non-perturbative 
corrections in the supersymmetric gauge theory. 
So we can expect that the volume has another side view as 
the non-perturbative corrections in the supersymmetric 
gauge theory.
 
We have concentrated in the case of vortex 
to evaluate the volume of the moduli space. 
Our method of topological field theory should be extendable 
to various problems in solitons and (supersymmetric) gauge 
theories. 
Our localization method can be applicable to other types 
of solitons like domain-walls and monopoles, 
which have codimension one and three, respectively.
Once we have the BPS equations for these solitons, 
we should be able to construct an appropriate ``action'' 
and BRST symmetry from the constraints. 
The localization of the path integral will give the volume. 
We can obtain the thermodynamical properties of these 
solitons from the volume. 
These BPS solitons also play a very important role in 
supersymmetric gauge theories. 
We expect that  
the evaluation of the volume shed light on 
the non-perturbative aspects of the supersymmetric 
gauge theories like dualities and confinement.

\section*{Acknowledgements}
One of the authors (K.O.) would like to thank M.~Nitta, K.~Ohashi and Y.~Yoshida
for useful discussions and comments.
One of the authors (N.S.) thanks Nick Manton, Norman Rink, 
Makoto Sakamoto, and David Tong for a useful discussion.
This work is supported in part by Grant-in-Aid for
Scientific Research from the Ministry of Education,
Culture, Sports, Science and Technology, Japan No.19740120 (K.O.),
No.21540279 (N.S.) and No.21244036 (N.S.), and by Japan Society
for the Promotion of Science $\langle$JSPS$\rangle$ and Academy of Sciences of
the Czech Republic $\langle$ASCR$\rangle$ under the Japan - Czech Republic
Research Cooperative Program (N.S.).

\appendix
\section{Volume of Grassmannian} 

The complex Grassmann manifold (Grassmannian) $G_{N_c,N_f}$ is defined by a homogeneous coset space of the unitary groups
\be
G_{N_c,N_f} = \frac{U(N_f)}{U(N_c)\times U(N_f-N_c)}.
\ee
The volume of $G_{N_c,N_f}$ is expressed by the volume of each unitary groups \cite{Fujii}
\be
\Vol(G_{N_c,N_f}) = \frac{\Vol(U(N_f))}{\Vol(U(N_c))\times \Vol(U(N_f-N_c))}.
\ee
The volume of $U(N)$ group is given by \cite{Macdonald,Ooguri:2002gx}
\be
\Vol(U(N))=\frac{(2\pi)^{N(N+1)/2}}{G(N+1)},
\ee
where $G(z)$ is the Barnes G-function which satisfies
\be
G(z+1)=\Gamma(z)G(z),
\ee
that is 
\be
G(N+1) = \prod_{i=1}^N \Gamma(i) =  \prod_{i=1}^{N}(i-1)! \quad \text{for }N\in \Z_{>0}.
\ee
Using this formula, we find the volume of the Grassmannian $G_{N_c,N_f}$
\bea
\Vol(G_{N_c,N_f})
&=& (2\pi)^{N_c(N_f-N_c)}
\frac{ \prod_{i=1}^{N_c}(i-1)! \times  \prod_{i=1}^{N_f-N_c}(i-1)! }{ \prod_{i=1}^{N_f}(i-1)! }\nn\\
&=&(2\pi)^{N_c(N_f-N_c)} \prod_{i=1}^{N_c}\frac{(i-1)!}{(N_f-i)!}
.
\eea
For example, the volume of complex projective space
$
\C P^{N_f-1} \simeq G_{1,N_f}
$
is given by
\be
\Vol(\C P^{N_f-1}) = \frac{(2\pi)^{N_f-1}}{(N_f-1)!}
\ee
in this normalization.



\begin{thebibliography}{99}
  
\bibitem{Manton:2004tk}
  N.~S.~Manton and P.~Sutcliffe,
  ``Topological solitons,''
{\it  Cambridge, UK: Univ. Pr. (2004) 493 p}.

\bibitem{Manton:1981mp}
  N.~S.~Manton,
  Phys.\ Lett.\  B {\bf 110} (1982) 54.

\bibitem{Samols:1991ne}
  T.~M.~Samols,
  Commun.\ Math.\ Phys.\  {\bf 145}, 149 (1992).

\bibitem{Manton:1993tt}
  N.~S.~Manton,
  Nucl.\ Phys.\  B {\bf 400} (1993) 624.

\bibitem{Shah:1993us}
  P.~A.~Shah and N.~S.~Manton,
  J.\ Math.\ Phys.\  {\bf 35} (1994) 1171
  [arXiv:hep-th/9307165].

\bibitem{Nekrasov:2002qd}
  N.~A.~Nekrasov,
  Adv.\ Theor.\ Math.\ Phys.\  {\bf 7} (2004) 831
  [arXiv:hep-th/0206161].


\bibitem{Taubes:1979tm}
  C.~H.~Taubes,
  Commun.\ Math.\ Phys.\  {\bf 72}, 277 (1980).

\bibitem{Manton:1998kq}
  N.~S.~Manton and S.~M.~Nasir,
  Commun.\ Math.\ Phys.\  {\bf 199} (1999) 591
  [arXiv:hep-th/9807017].

\bibitem{Nasir:1998kt}
  S.~M.~Nasir,
  Phys.\ Lett.\  B {\bf 419} (1998) 253
  [arXiv:hep-th/9807020].

\bibitem{Manton:2002wb}
  N.~S.~Manton and J.~M.~Speight,
  Commun.\ Math.\ Phys.\  {\bf 236} (2003) 535
  [arXiv:hep-th/0205307].

\bibitem{Chen:2004xu}
  H.~Y.~Chen and N.~S.~Manton,
  J.\ Math.\ Phys.\  {\bf 46} (2005) 052305
  [arXiv:hep-th/0407011].

\bibitem{Romao:2005ph}
  N.~M.~Romao,
  J.\ Phys.\ A  {\bf 38} (2005) 9127
  [arXiv:hep-th/0503014].


\bibitem{ANO}
A. Abrikosov, Sov.\ Phys.\ JETP {\bf 32} (1957) 1442,\\
H. Nielsen and P. Olesen, Nucl.\ Phys.\ B {\bf 61} (1973) 45.

\bibitem{Bradlow:1990ir}
  S.~B.~Bradlow,
  Commun.\ Math.\ Phys.\  {\bf 135} (1990) 1.

\bibitem{Manton:2010sa}
  N.~S.~Manton and N.~M.~Romao,
  arXiv:1010.0644 [hep-th].

\bibitem{Manton:2009ja}
  N.~S.~Manton and N.~A.~Rink,
  J.\ Phys.\ A  {\bf 43} (2010) 434024
  [arXiv:0912.2058 [hep-th]]; arXiv:1012.3014 [hep-th].




\bibitem{Hanany:2003hp}
  A.~Hanany and D.~Tong,
  JHEP {\bf 0307} (2003) 037
  [arXiv:hep-th/0306150];
%
  A.~Hanany and D.~Tong,
  JHEP {\bf 0404} (2004) 066
  [arXiv:hep-th/0403158].


\bibitem{Eto:2005yh}
  M.~Eto, Y.~Isozumi, M.~Nitta, K.~Ohashi and N.~Sakai,
  Phys.\ Rev.\ Lett.\  {\bf 96} (2006) 161601
  [arXiv:hep-th/0511088].

\bibitem{Eto:2006mz}
  M.~Eto, T.~Fujimori, Y.~Isozumi, M.~Nitta, K.~Ohashi, K.~Ohta and N.~Sakai,
  Phys.\ Rev.\ D {\bf 73} (2006) 085008
  [arXiv:hep-th/0601181].

\bibitem{Eto:2006pg}
  M.~Eto, Y.~Isozumi, M.~Nitta, K.~Ohashi and N.~Sakai,
  J.\ Phys.\ A  {\bf 39} (2006) R315
  [arXiv:hep-th/0602170].

\bibitem{Eto:2006uw}
  M.~Eto, Y.~Isozumi, M.~Nitta, K.~Ohashi and N.~Sakai,
  Phys.\ Rev.\  D {\bf 73}, 125008 (2006)
  [arXiv:hep-th/0602289].

\bibitem{Eto:2007yv}
  M.~Eto, J~Evslin, K.~Konishi, G.~Marmorini, M.~Nitta, K.~Ohashi, 
W.~Vinci and N.~Yokoi,
  Phys.\ Rev.\  D {\bf 76} (2007) 105002
  [arXiv:0704.2218 [hep-th]].

\bibitem{Eto:2007aw}
  M.~Eto, T.~Fujimori, M.~Nitta, K.~Ohashi, K.~Ohta and N.~Sakai,
  Nucl.\ Phys.\  B {\bf 788} (2008) 120
  [arXiv:hep-th/0703197].

\bibitem{Baptista:2008ex}
  J.~M.~Baptista,
  Commun.\ Math.\ Phys.\  {\bf 291} (2009) 799
  [arXiv:0810.3220 [hep-th]].

\bibitem{Eto:2009wq}
  M.~Eto, T.~Fujimori, T.~Nagashima, M.~Nitta, K.~Ohashi and N.~Sakai,
  Phys.\ Lett.\  B {\bf 678}, 254 (2009)
  [arXiv:0903.1518 [hep-th]].

\bibitem{Manton:2010wu}
  N.~S.~Manton and N.~Sakai,
  Phys.\ Lett.\  B {\bf 687} (2010) 395
  [arXiv:1001.5236 [hep-th]].

%
%
%

\bibitem{Fujimori:2010fk}
  T.~Fujimori, G.~Marmorini, M.~Nitta, K.~Ohashi and N.~Sakai,
  Phys.\ Rev.\  D {\bf 82} (2010) 065005
  [arXiv:1002.4580 [hep-th]].




\bibitem{Moore:1997dj}
  G.~W.~Moore, N.~Nekrasov and S.~Shatashvili,
  Commun.\ Math.\ Phys.\  {\bf 209} (2000) 97
  [arXiv:hep-th/9712241].

\bibitem{Gerasimov:2006zt}
  A.~A.~Gerasimov and S.~L.~Shatashvili,
  Commun.\ Math.\ Phys.\  {\bf 277} (2008) 323
  [arXiv:hep-th/0609024].
  
\bibitem{Witten:1992xu}
  E.~Witten,
  J.\ Geom.\ Phys.\  {\bf 9} (1992) 303
  [arXiv:hep-th/9204083].

\bibitem{Blau:1993tv}
 M.~Blau and G.~Thompson,
 Nucl.\ Phys.\  B {\bf 408} (1993) 345,
 hep-th/9305010; hep-th/9310144.



\bibitem{Lee:2006ys}
  K.~M.~Lee and H.~U.~Yee,
  JHEP {\bf 0703} (2007) 012
  [arXiv:hep-th/0605214],\\
  H.~U.~Yee,
  Nucl.\ Phys.\  B {\bf 774} (2007) 232
  [arXiv:hep-th/0612002].


\bibitem{Fujii}
K.~Fujii,
J. Appl. Math. {\bf 2} (2002), 371-405.

\bibitem{Macdonald}
I.~G.~Macdonald,
Inventiones Math.\ {\bf 56} (1980) 93-95.

\bibitem{Ooguri:2002gx}
  H.~Ooguri and C.~Vafa,
  Nucl.\ Phys.\  B {\bf 641} (2002) 3
  [arXiv:hep-th/0205297].


\bibitem{Nakahara:1990th}
  M.~Nakahara,
  ``Geometry, topology and physics,''
{\it  Bristol, UK: Hilger (1990) 505 p. (Graduate student series in physics)}.

\bibitem{Hitchin:1986ea}
  N.~J.~Hitchin, A.~Karlhede, U.~Lindstrom and M.~Rocek,
  ``Hyperkahler Metrics and Supersymmetry,''
  Commun.\ Math.\ Phys.\ {\bf 108} (1987) 535.



%
%


\bibitem{Moore:1998et}
  G.~W.~Moore, N.~Nekrasov and S.~Shatashvili,
  Commun.\ Math.\ Phys.\  {\bf 209} (2000) 77
  [arXiv:hep-th/9803265].


\bibitem{Shadchin:2006yz}
  S.~Shadchin,
  JHEP {\bf 0708} (2007) 052
  [arXiv:hep-th/0611278].


\bibitem{Ohta:2007ji}
  K.~Ohta,
  arXiv:0710.4011 [hep-th].

\bibitem{Dimofte:2010tz}
  T.~Dimofte, S.~Gukov and L.~Hollands,
  arXiv:1006.0977 [hep-th].

\bibitem{Awata:2010bz}
  H.~Awata, H.~Fuji, H.~Kanno, M.~Manabe and Y.~Yamada,
  arXiv:1008.0574 [hep-th].

\bibitem{Yoshida:2011au}
  Y.~Yoshida,
  arXiv:1101.0872 [hep-th].

\bibitem{Bonelli:2011fq}
  G.~Bonelli, A.~Tanzini and J.~Zhao,
  arXiv:1102.0184 [hep-th].

\bibitem{Nakajima:1999}
H.~Nakajima,
``Lectures on Hilbert Schemes of Points on Surfaces,''
{\it American Math- ematical Society,University Lectures Series v.18 (1999)}.

\bibitem{Bruzzo:2002xf}
  U.~Bruzzo, F.~Fucito, J.~F.~Morales and A.~Tanzini,
  JHEP {\bf 0305} (2003) 054
  [arXiv:hep-th/0211108].



\end{thebibliography}
\end{document}